\documentclass[usenatbib]{mn2e}

\usepackage[english]{babel}
\usepackage[T1]{fontenc}
\usepackage{amsmath}
\usepackage{bbold}
\usepackage{graphicx}
\usepackage{amsfonts}
\usepackage{amssymb}
\usepackage{latexsym}

\newcommand{\be}{\begin{equation}}
\newcommand{\ee}{\end{equation}}

\newcommand{\mnras}{{MNRAS }}
\newcommand{\aap}{{A\&A}}
\newcommand{\pasp}{{PASP}}
\newcommand{\pasj}{{PASJ}}
\newcommand{\aj}{{AJ}}
\newcommand{\apj}{{ApJ}}
\newcommand{\apjs}{{ApJS}}

\usepackage{color}

\begin{document}

\title[Measuring large-scale structure with quasars]{Measuring large-scale structure with quasars in narrow-band filter surveys}

\author[L.~R.~Abramo et al.]
{L.~Raul Abramo, $^{1,2,3}$ Michael~A.~Strauss, $^1$ Marcos~Lima, $^{2,3,4}$
\newauthor
%I.~Gonzalez-Serrano, $^{?}$ 
Carlos~Hern\' andez-Monteagudo, $^5$ Ruth~Lazkoz, $^6$ Mariano~Moles, $^5$
\newauthor
Claudia ~Mendes~de~Oliveira, $^4$ Irene~Sendra $^6$, Laerte~Sodr\'e Jr. $^4$
\newauthor 
and Thaisa Storchi-Bergmann $^7$\\
$^1$ Department of Astrophysical Sciences, Princeton University, 
Peyton Hall, Princeton, NJ 08544 \\
$^2$ Department of Physics \& Astronomy, University of Pennsylvania, 
Philadelphia, PA, 19104 \\
$^3$ Dep. de F\'{\i}sica Matem\'atica,
Instituto de F\'{\i}sica, Universidade de S\~ao Paulo, 
CP 66318, CEP 05314-970 São Paulo, SP, Brazil \\
$^4$ Departamento de Astronomia, IAG, Universidade de S\~ao Paulo, 
Rua do Mat\~ao 1226, CEP 05508-090 S\~ao Paulo, SP, Brazil \\
$^5$ Centro de Estudios de F\' {\i}sica del
Cosmos de Arag\' on (CEFCA), Plaza San Juan 1, planta 2, E-44001,
Teruel, Spain \\
$^6$ Fisika Teorikoa, Zientzia eta Teknologia Fakultatea, Euskal Herriko Unibertsitatea, 644 Posta Kutxatila, 48080 Bilbao, Spain \\
$^7$ Instituto de F\'{\i}sica, Universidade Federal do Rio Grande do
Sul, Av. Bento Gon\c{c}alves 9500, 91501-970, Porto Alegre, 
RS, Brazil}

\maketitle

\begin{abstract}
We show that a large-area imaging survey using narrow-band 
filters could detect quasars in
sufficiently high number densities, and with more than sufficient
accuracy in their photometric redshifts, to turn them into suitable 
tracers of large-scale structure.
If a narrow-band optical survey can detect objects as faint as 
$i = 23$, it could reach volumetric number densities as high as
$10^{-4} \, h^3$ Mpc$^{-3}$ (comoving) at $z \sim 1.5$.
Such a catalog would lead to precision measurements of the power 
spectrum up to $z \sim 3-4$.
We also show that it is possible to employ quasars to measure 
baryon acoustic oscillations at high redshifts, where the
uncertainties from redshift distortions and nonlinearities are 
much smaller than at $z \lesssim 1$. 
As a concrete example we study the future impact of J-PAS, 
which is a narrow-band imaging survey in the optical over 1/5 
of the unobscured sky with $42$ filters of $\sim 100$  {\AA} 
full-width at half-maximum.
We show that J-PAS will be able to take advantage of the broad 
emission lines of quasars to deliver excellent photometric redshifts, 
$\sigma_{z} \simeq 0.002 \, (1+z) $, for millions of objects.
% These excellent photometric redshifts would allow not only
% a complete characterization of the quasar luminosity function, bias,
% clustering and constraints on duty cycles, but would also lead to
% the reconstruction of the power spectrum at several
% high-redshift intervals, and even a (low-statistics) estimate 
% of radial BAOs.
\end{abstract}

%\pacs{98.80.C9; 98.80.-k; 98.65.Cw}

%%%%%%%%%%%%%%%%%
% Next section
%%%%%%%%%%%%%%%%%

\section{Introduction}

Quasars are among the most luminous objects in the Universe.
They are believed to be powered by the accretion disks of
giant black holes that lie at the centers of galaxies
[\citet{salpeter_accretion_1964,zeldovich_mass_1965,lynden-bell_galactic_1969}], 
and the extreme environments of those disks are responsible for emitting the 
``non-stellar continuum'' and the broad emission lines that characterize the 
spectral energy distributions (SEDs) of quasars and most other 
types of Active Galactic Nuclei (AGNs).

However, even though all galaxy bulges in the local Universe seem to host
supermassive black holes in their centers [\citet{kormendy_inward_1995}], the 
duty cycle of quasars is much smaller than the age 
of the Universe [\citet{richstone_supermassive_1998}].
This means that at any given time the number 
density of quasars is small compared to that of galaxies.

As a consequence, galaxies have been the preferred tracers of large-scale
structure in the Universe: their high densities and relatively high luminosities
allow astronomers to compile large samples, distributed across vast volumes. 
Both spectroscopic [see, e.g., \citet{cole_2df_2005,York:2000gk};
BOSS\footnote{http://cosmology.lbl.gov/BOSS/}]
and broad-band (e.g., $ugriz$) photometric surveys 
[\citet{scoville_cosmic_2007,adelman-mccarthy_sdss_2008,adelman-mccarthy_sixth_2008}]
have been used with remarkable success to study the distribution 
of galaxies, particularly so for the subset of 
luminous red galaxies (LRGs), for which it is possible to obtain 
relatively good photometric redshifts (photo-z's) 
[$\sigma_z \sim 0.01 \, (1+z)$] even with broad-band filter photometry
[\citet{bolzonella_photometric_2000,benitez_bayesian_2000,Firth_2003,padmanabhan_calibrating_2005,Padmanabhan:2006ku,abdalla_comparison_2008,abdalla_photometric_2008,abdalla_predicting_2008}].
From a purely statistical perspective, 
photometric surveys have the advantage of larger 
volumes and densities than spectroscopic surveys 
-- albeit with diminished spatial resolution in the radial direction, 
which can be a limiting factor for some applications, in particular baryon 
acoustic oscillations (BAOs) [\citet{blake_cosmology_2005}].

Most ongoing wide-area surveys choose one of the
parallel strategies of imaging 
[e.g., \citet{Abbott:2005bi};
PAN-STARRS\footnote{Pan-STARRS technical summary, http://pan-starrs.ifa.hawaii.edu/public/};\citet{LSST:2009pq}] or
multi-object spectroscopy (e.g., BOSS), and future instruments 
will probably continue following these trends, since spectroscopic
surveys need wide, deep imaging for target selection, and imaging
surveys need large spectroscopic samples as calibration sets.

However, whereas LRGs possess a signature spectral feature (the 
so-called $\lambda_{\rm rest} \sim$ 4,000 {\AA}  break), which 
translates into fairly good photo-z's with $ugriz$ imaging, the SEDs of quasars  
observed by broad-band filters only show a similar 
feature (the Ly-$\alpha$ line) at $\lambda_{\rm rest} \sim$ 1,200 \AA, which
makes them UV-dropout objects.
The segregation of quasars from stars and unresolved galaxies
in color-color and color-magnitude diagrams has allowed the 
construction of a high-purity catalog of $\sim 1.2 \times 10^6$ 
photometrically selected quasars in the SDSS [\citet{richards_efficient_2008}], 
and the (broad-band) photometric redshifts of $z< 2.2$ objects in that catalog 
can be estimated by the passage of the emission lines from one filter
to the next [\citet{richards_photometric_2001}].
More recently, \citet{2009ApJ...690.1250S} showed that a 
combination of broad-band and medium-band filters reduced the photo-z 
errors of the XMM-COSMOS sources down to $\sigma_z/(1+z) \sim 0.01$ (median).

The SDSS spectroscopic catalog of quasars  
[\citet{schneider_sloan_2003,schneider_sloan_2007,schneider_sloan_2010}] 
is five times bigger than
previous samples [\citet{croom_2df_2004}], but includes only $\sim 10$\% of 
the total number of good candidates in the photometric sample. Furthermore, 
that catalog is limited to relatively bright objects, with apparent magnitudes 
$i \lesssim 19.1$ at $z< 3.0$, and $i<20.2$ for objects with $z> 3.0$.
Despite the sparseness of the SDSS spectroscopic catalog of quasars 
(the comoving number density of objects in that catalog peaks at 
$\lesssim 10^{-6}$ Mpc$^{-3}$ around $z \sim 1$), it has been successfully 
employed in several measurements of large-scale structure -- 
see, e.g., \citet{porciani_cosmic_2004,da_angela_2df_2005}, 
which used the 2QZ survey [\citet{croom_2005,croom_2df-sdss_2009}]
for the first modern applications of quasars in a cosmological context; 
see \citet{shen_clustering_2007,ross_clustering_2009-1} for the cosmological
impact of the SDSS quasar survey;
and \citet{padmanabhan_real-space_2008}, which 
cross-correlated quasars with the SDSS photometric catalog of LRGs.
One can also use quasars as a backlight to illuminate the intervening 
distribution of neutral H, which can then be used to compute the mass power spectrum 
[\citet{croft_recovery_1997,seljak_cosmological_2005}].
%The Baryon Oscillation Spectroscopic 
%Survey (BOSS, \citet{BOSS}), in particular, is concentrating on taking spectra 
%of galaxies rather than quasars, which means that it will not yield
%a sufficiently dense catalog of quasars to allow their use as tracers of
%large-scale structure.

The broad emission lines of type-I quasars [\citet{vanden_berk_composite_2001}],
which are a manifestation of the extremely high velocities
of the gas in the environments of supermassive black holes, are 
ideal features with which to obtain photo-z's, if
only the filters were narrow enough ($\Delta \lambda \lesssim 400$ 
\AA) to capture those features. 
%% ADDED COMBO-17 REFERENCE HERE
In fact, \citet{Wolf_COMBO17_q} have produced a sample of a few 
hundred quasars using a combination of broad-band and narrow-band filters 
(the COMBO-17 survey), obtaining a photo-z accuracy of approximately 3\%
-- basically the same accuracy that was for their catalog of galaxies 
[\citet{Wolf_COMBO17_g}].
%%
%Since the effective {\it {\'e}tendue} 
%(area of the field of view times the area of the mirror)
%attainable with an imaging survey is typically much higher than that  
%with a comparable fiber-based multi-object spectroscopic survey, 
Acquiring a sufficiently large number of quasars in an existing narrow-band galaxy 
survey would be both feasible and it would bear zero marginal cost 
on the survey budgets.

Fortunately, a range of science cases that hinge on large volumes and 
good spectral resolution, in particular galaxy surveys with the goal of 
measuring BAOs [\citet{Peebles:1970ag,sunyaev_interaction_1970,Bond:1984fp,holtzman_microwave_1989,Hu:1995en}],
both in the angular [\citet{BAO,tegmark_cosmological_2006,blake_cosmological_2007,Padmanabhan:2006ku,
Percival:2007yw}] and in the radial directions
[\citet{eisenstein_power_1997,eisenstein_deprojecting_2003,2003ApJ...594..665B,Seo:2003pu,angulo_detectability_2007,Seo:2007ns}]
has stimulated astronomers to construct new 
instruments. They should be not only capable of detecting huge numbers of 
galaxies, but also of measuring much more precisely the photometric 
redshifts for these galaxies -- and that means either 
low-resolution spectroscopy,
or filters narrower than the $u g r i z$ system.

Presently there are a few instruments which can be characterized either as
narrow-band imaging surveys, or low-resolution multi-object spectroscopy surveys: 
the Alhambra survey [\citet{moles_alhambra_2008}], 
PRIMUS [\citet{2008PhDT.........8C}],  
HETDEX\footnote{http://hetdex.org/hetdex/scientific\_papers.php} 
and the PAU survey\footnote{http://www.pausurvey.org}.
The Alhambra survey uses the LAICA camera on 
the 3.5 m Calar Alto telescope, and is mapping
4 deg$^2$ between 3,500 {\AA} and 9,700 {\AA}, 
using a set of 20 filters equally spaced in the optical 
plus $JHK$ broad filters in the NIR. 
PRIMUS takes low-resolution spectra of selected objects
with a prism and slit mask built for the IMACS instrument at the 6.5 m 
Magellan/Baade telescope. PRIMUS has already mapped $\sim$ 10 deg$^2$ of 
the sky down to a depth of 23.5, and has extracted redshifts of $\sim 3 \times 10^5$ 
galaxies up to $z=1$, with a photo-z accuracy of order 1\% [\citet{PRIMUS}].
HETDEX is a large-field of view, integral
field unit spectrograph to be mounted on the 10 m Hobby-Eberly telescope 
that will map 420 deg$^2$ with filling factor
of 1/7 and an effective spectral resolution of 6.4 {\AA} between 3,500
and 5,500 {\AA}. The PAU survey will use 40 narrow-band filters and five 
broadband filters mounted on a new camera on the William Herschel 
Telescope to observe 100-200 deg$^2$ down to a magnitude $i_{AB}\sim23$.
All these surveys will detect large numbers of 
intermediate- to high-redshift objects (including AGNs), 
and by their nature will provide very dense, extremely
complete datasets.

\begin{figure}
%\vspace{0.5cm}
\includegraphics[width=80mm]{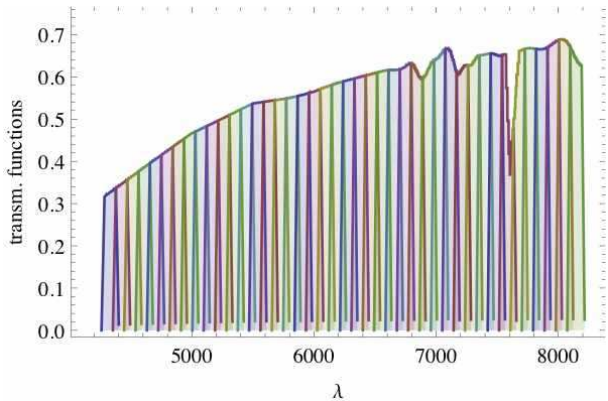}
\caption{Throughputs of the original J-PAS filter system, assuming an airmass of 1.2, two aluminum reflections and the quantum efficiency of the LBNL CCDs (N. Ben\'{\i}tez, private communication). The 42 narrow-band filters are spaced by 93 {\AA}, with 118 {\AA} FWHM, and span the interval between 4,250 {\AA} and 8,200 {\AA}. The final filter system for J-PAS is still under review, and may present small deviations from the original filter set of 
Ben\'{\i}itez {\it et al.} (2009) -- see Ben\'{\i}itez {\it et al.} (2011), to appear. We have 
checked that the results presented in this paper are basically insensitive to these small 
variations.}
\label{Fig:filters}
\end{figure}

\begin{figure}
\vspace{0.5cm}
%\includegraphics[width=16cm]{Vanden_Berk_template}
%\\
\includegraphics[width=75mm]{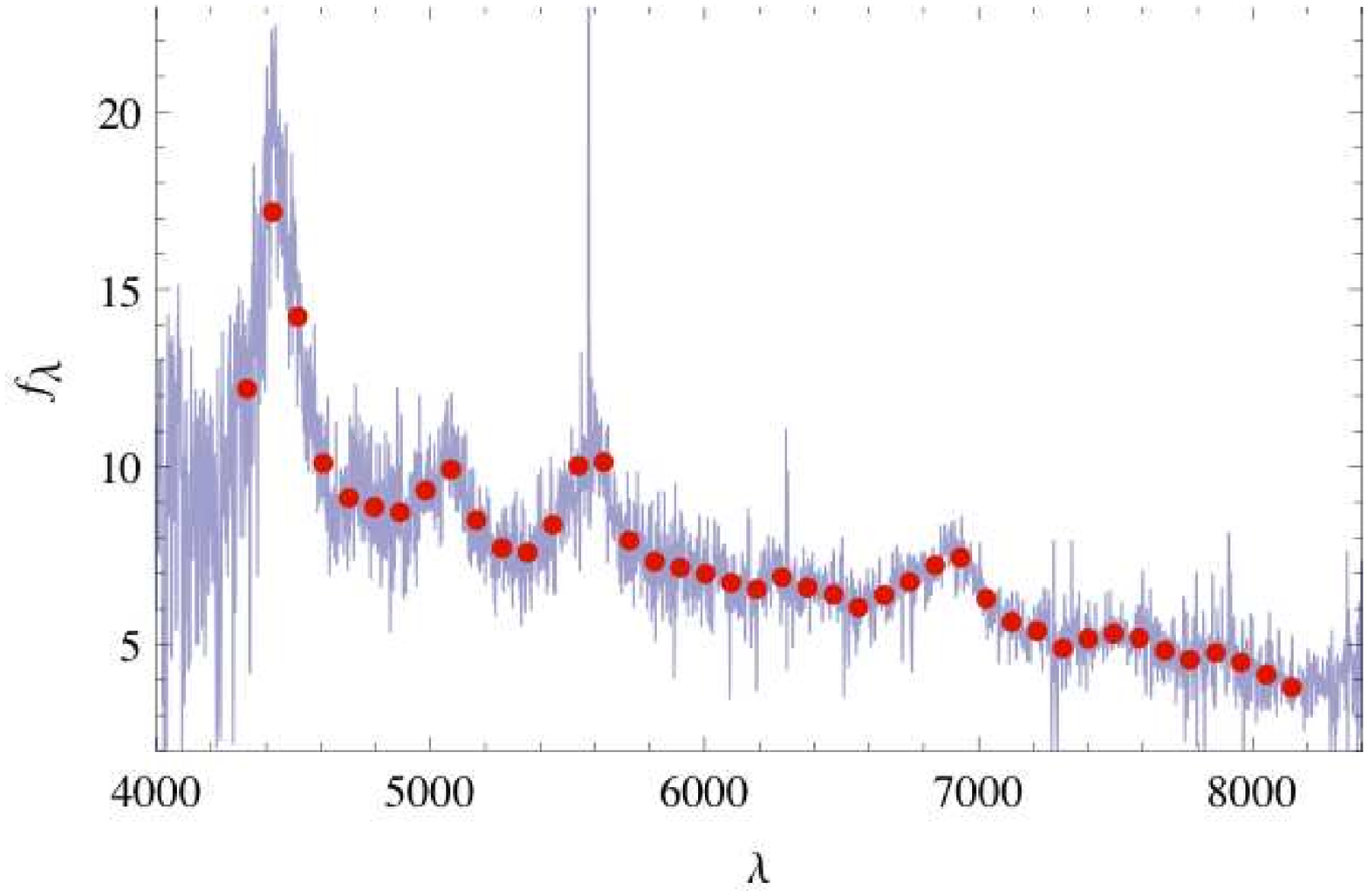}
\includegraphics[width=75mm]{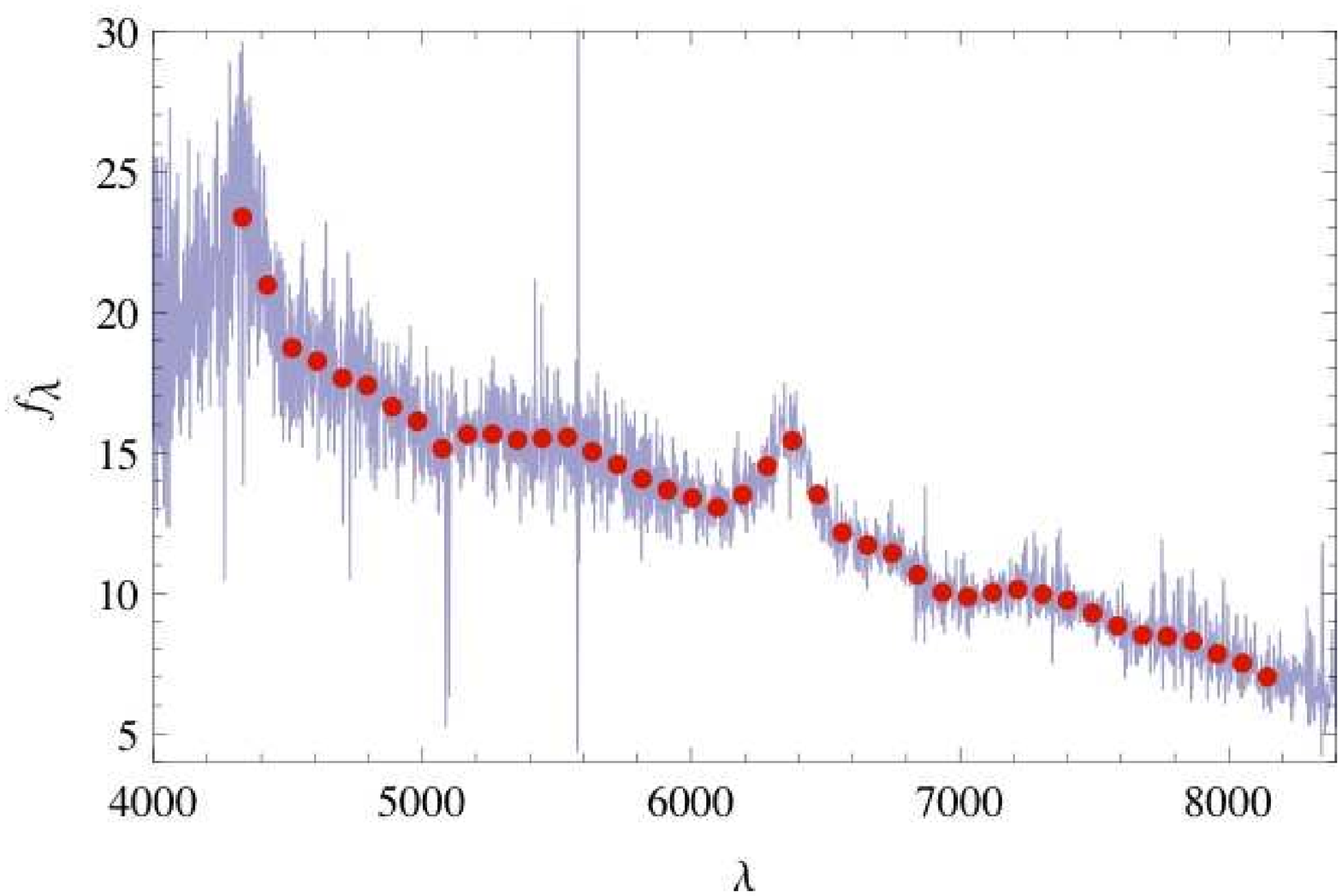}
\includegraphics[width=75mm]{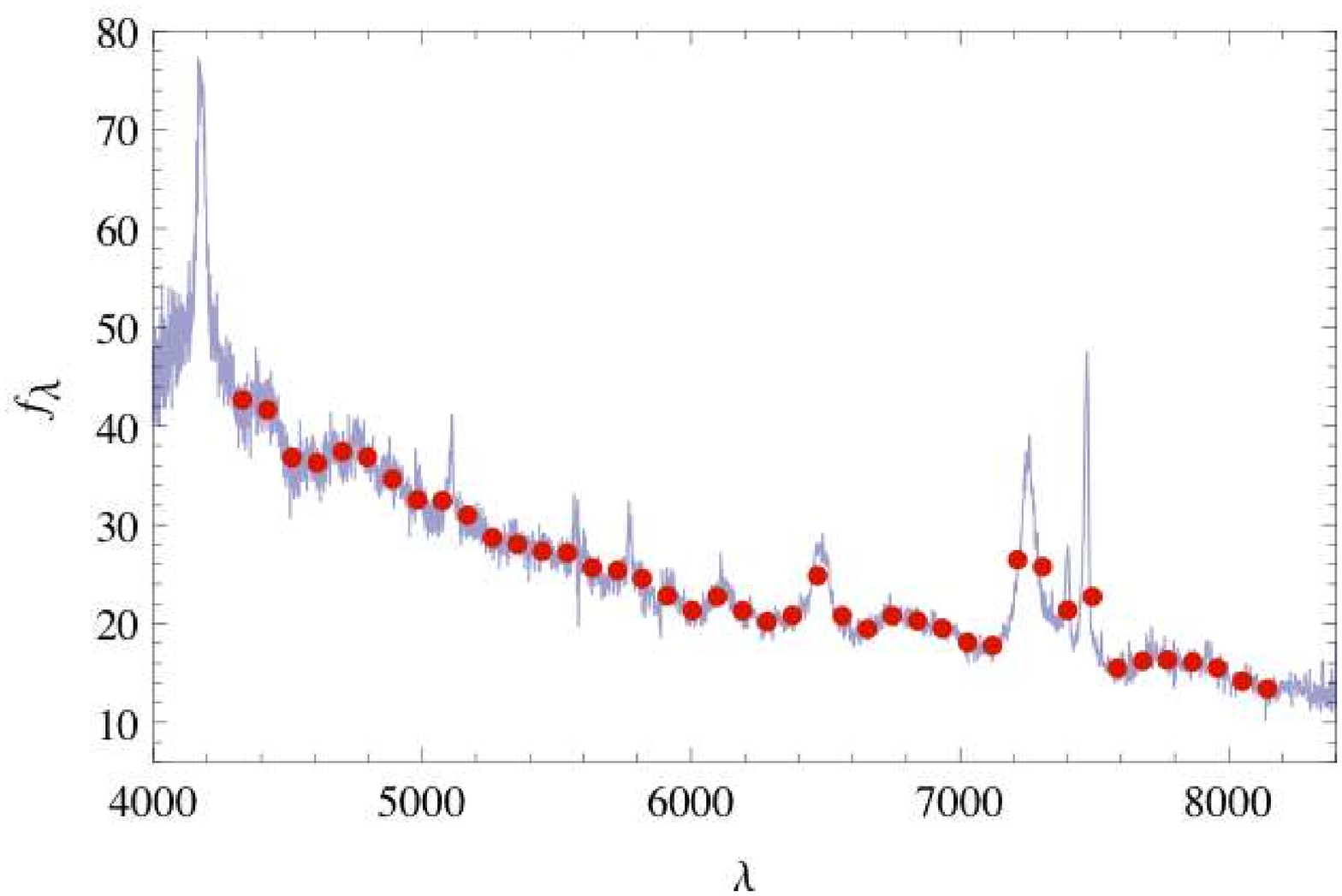}
\caption{Three SDSS quasars as they would be observed by the
filter system of Fig. \ref{Fig:filters}. The SDSS objects are, from top to bottom:
J000143.41-152021.4 ($z=2.638$),
J001138.43-104458.2 (at $z=1.271$), and
J002019.22-110609.2 ($z=0.492$). 
The light (blue in color version) curve indicates the flux (in units of $10^{-17}$ erg/s/cm$^2$/\AA) 
in spectral bins of the original SDSS spectra; the large (red) dots denote the corresponding 
fluxes (normalized by the filter throughput) for the J-PAS narrow-band filters. 
Some emission lines can be seen in the
photometric data: Ly-$\alpha$, Si IV, C IV and C III] for the spectrum on top;
C III] and Mg II for the quasar in the central panel; and Mg II, H$\gamma$ and 
H$\beta$ (together with the [O III] doublet) for the spectrum on the bottom.
}
\label{Fig:quasars}
\end{figure}

Another instrument which plans to make a wide-area spectrophotometric
map of the sky is the {\it Javalambre Physics of the Accelerating Universe
Astrophysical Survey} (J-PAS). The instrument [see \citet{Benitez:2008fs}] will 
consist of two telescopes, of 2.5 m (T250) and 0.8 m (T80) apertures, which are 
being built at Sierra de Javalambre, in mainland Spain (40$^{\rm o}$ N)
[\citet{moles_site_2010}]. 
A dedicated 1.2 Gpixel survey camera with a field of view of 7 deg$^2$ 
(5 deg$^2$ effective)
will be mounted on the focal plane of the T250 telescope, while the T80 
telescope will be used mainly for photometric calibration.
The survey (which is fully funded through a Spain-Brazil collaboration) 
is planned to take 4-5 years and is expected
to map between 8,000 and 9,000 deg$^2$ to a 5$\sigma$ magnitude depth 
for point sources equivalent to $i_{\rm AB} \sim 23$ 
($i \sim 23.3$)  over an aperture of 2 arcsec$^2$. The filter system 
of the J-PAS instrument, as originally described in \citet{Benitez:2008fs}, 
consists of 42 contiguous narrow-band filters of 118 {\AA} 
FWHM spanning the range from 4,300 {\AA}  to 8,150 {\AA}  -- see Fig. 
\ref{Fig:filters}. 
This set of filters was designed to extract photo-z's
of LRGs with (rms) accuracy as good as 
$\sigma_z \simeq 0.003 \, (1+z)$.
Of course, this filter configuration is also ideal to detect and
extract photo-z's of type-I quasars -- see Fig. \ref{Fig:quasars}.

%\begin{figure}

In this paper we show that a narrow-band imaging survey such
as J-PAS will detect quasars in sufficiently high numbers
($\sim 2. \times 10^6$ up to $z \simeq 5$), 
and with more than sufficient redshift accuracy, to make 
precision measurements of the power spectrum. In particular, 
these observations will yield a high-redshift measurement of BAOs, 
at an epoch where redshift distortions and 
nonlinearities are much less of a nuisance than in the local
Universe. This huge dataset may also allow precision 
measurements of the quasar luminosity function [\citet{2007ApJ...654..731H}], 
clustering and bias [\citet{shen_clustering_2007,ross_clustering_2009-1,
shen_catalog_2010}], 
% as well as redshift distortions [\citet{calvao_probing_2002}] 
as well as limits on the quasar duty cycle [\citet{martini_quasar_2001}].

%% CHANGED PARAGRAPH
This paper is organized as follows: 
In Section II we show how narrow 
($\sim 100$  {\AA} bandwidth) filters can be used to extract redshifts 
of quasars with high efficiency and accuracy. We compare two 
photo-z methods: empirical template fitting, and the training set method.
%% added
Still in Section II, we study the issues of completeness and contamination.
In Section III we compute the
expected number of quasars in a flux-limited narrow-band imaging 
survey, and derive the uncertainties in the power spectrum that can 
be achieved with that catalog. 
Our fiducial cosmological model is a flat $\Lambda$CDM Universe 
with $h=0.72$ and $\Omega_m=0.25$, and all distances are comoving, 
unless explicitly noted.

As we were finalizing this work, a closely related preprint, 
\citet{2011arXiv1108.1198S}, came to our notice.
In that paper the authors analyze the SDSS, 2QZ and 2SLAQ quasar 
catalogs in search of the BAO features -- see also  
\citet{2005PASJ...57..529Y} for a previous attempt using only the SDSS
quasars. Although Sawangwit {\it et al.} are unable to
make a detection of BAOs with these combined catalogs, they have forecast that a
spectroscopic survey with a quarter million quasars over 2000 deg$^2$ would be 
sufficient to detect the scale of BAOs with accuracy comparable to 
that presently made by LRGs -- but at a higher redshift. Their 
conclusions are consistent with what we have found in  
Section III of this paper.

\section{Photometric redshifts of quasars}

The idea of using the fluxes observed through multiple filters, 
instead of full-fledged spectra, to estimate the redshifts of astronomical objects, 
is almost five decades old [\citet{baum_photoelectric_1962}],
but only recently it has acquired greater relevance in connection
with photometric galaxy surveys
[\citet{Connolly_1995,bolzonella_photometric_2000,
benitez_bayesian_2000,blake_cosmology_2005,Firth_2003,budavari_unified_2009}]. 
In fact, many planned astrophysical surveys such as DES
[\citet{Abbott:2005bi}], Pan-STARRS
and the LSST [\citet{LSST:2009pq}]
are relying (or plan to rely) almost entirely on photometric 
redshifts (photo-z's) of galaxies for the bulk of their science cases.

Photometric redshift methods can be divided into two basic
groups: empirical template fitting methods, and training set methods -- 
see, however, \citet{budavari_unified_2009} for a unifying scheme. 
With template-based methods
[which may include spectral synthesis methods, e.g. 
\citet{bruzual_stellar_2003}] the photometric
fluxes are fitted (typically through a $\chi^2$) to some model, or
template, which has been properly redshifted, and the photometric
redshift ({\it photo-z}) is given by a maximum likelihood estimator (MLE).
In the training set approach, a large number 
of spectra is used to empirically calibrate a multidimensional mapping
between photometric fluxes and redshifts, without explicit modeling 
templates.

The performance of template fitting methods and of training set methods 
are similar when they are applied to broad-band photometric surveys
[\citet{budavari_unified_2009}]. 
In this paper we have taken both approaches, in order to compare their
performances specifically for the case of a narrow-band filter surveys of 
quasars.

\subsection{The spectroscopic sample of quasars}

\begin{figure*}
\vspace{0.5cm}
\includegraphics[width=80mm]{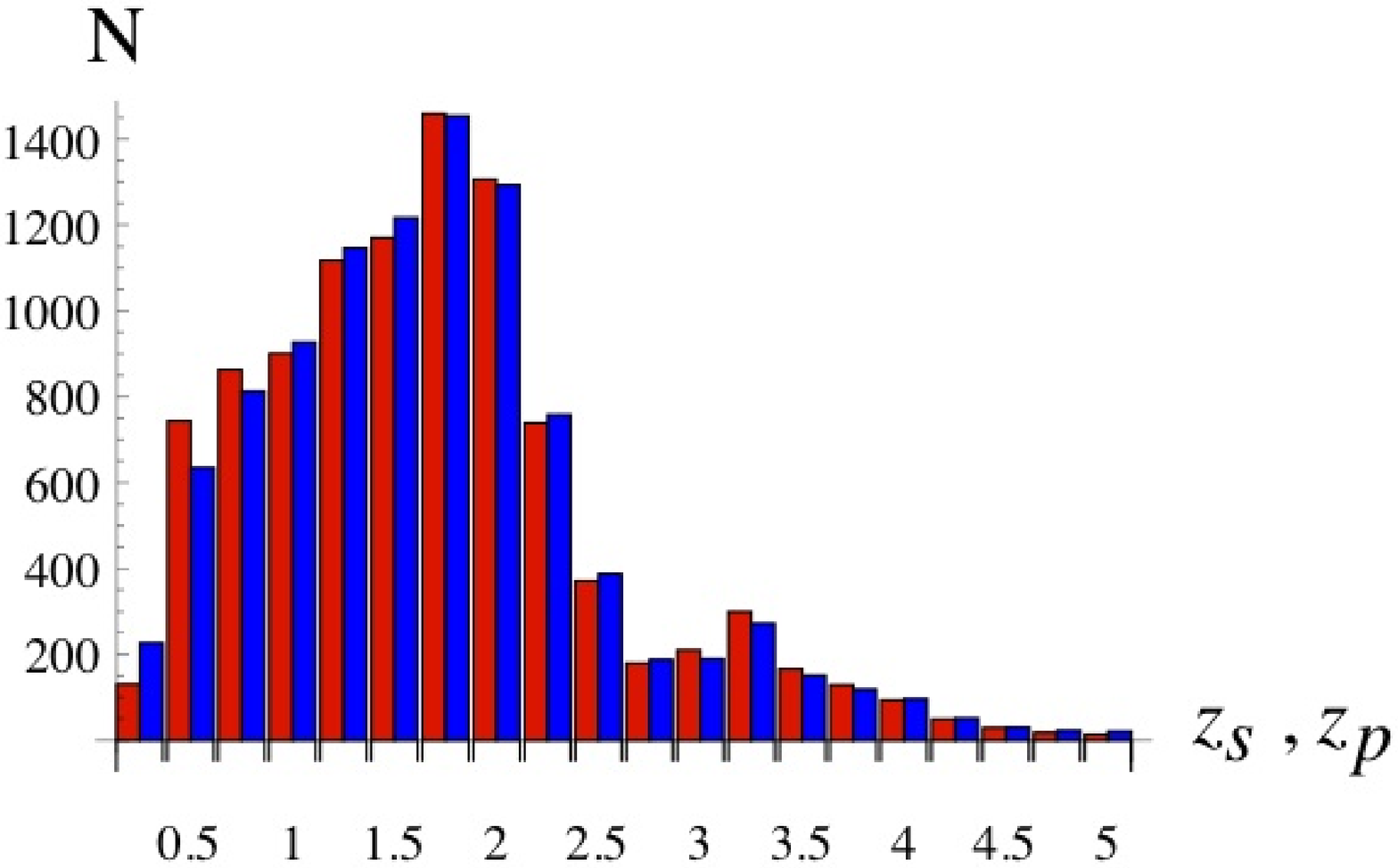}
\includegraphics[width=80mm]{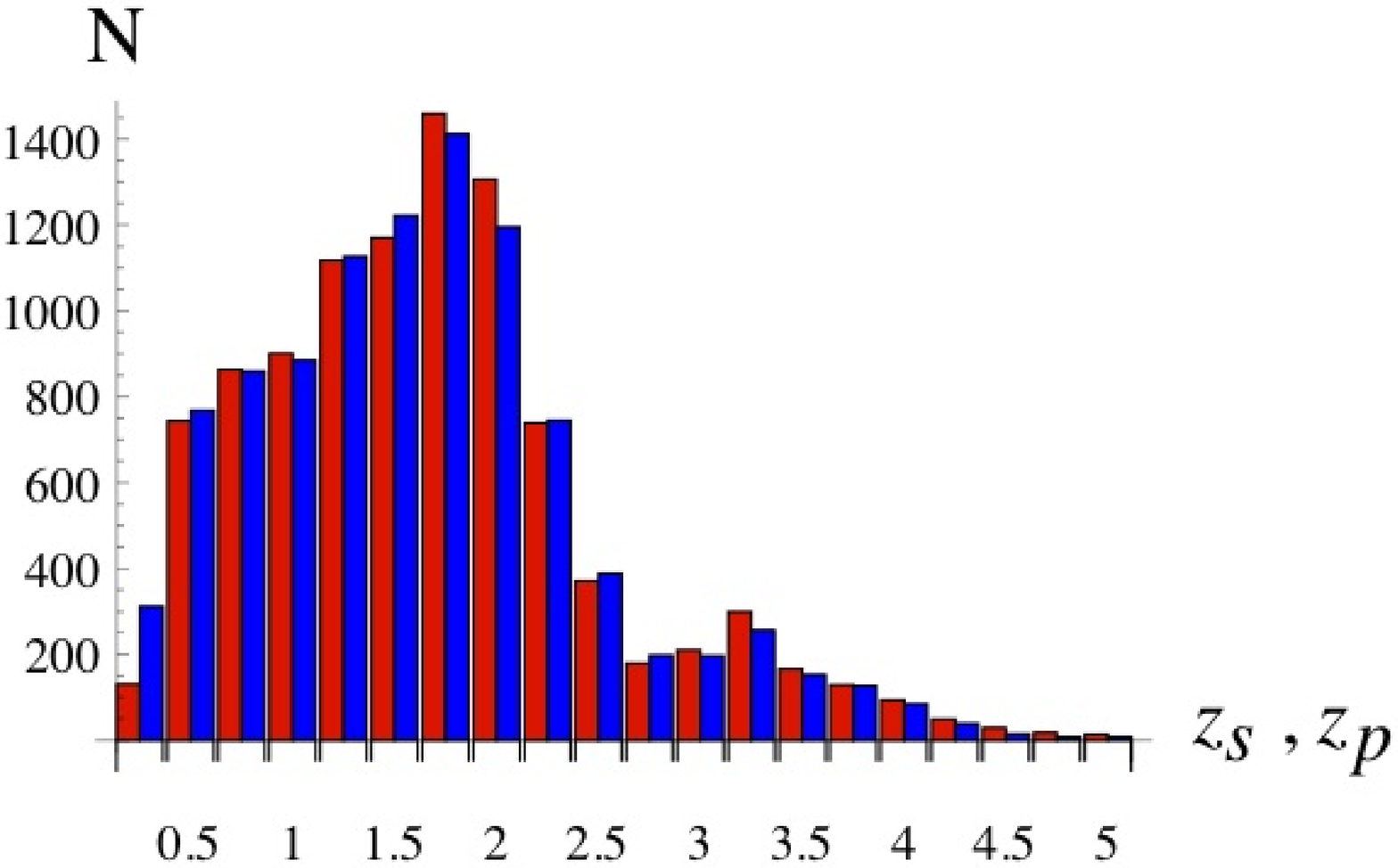}
\caption{
Redshift distribution of our full sample of quasars, in terms of their
spectroscopic redshifts $z_s$ (left bars, red in color version) and
their photometric redshifts $z_p$ obtained through the template fitting method
of Section \ref{Sec:Templates}
(right bars, blue in color version), in bins of $\Delta z = 0.25$.
Left panel: sub-sample
of SDSS quasars; right panel: simulated sample of fainter objects.
}
\label{Fig:SelFun}
\end{figure*}

We have randomly selected a sample of 10,000 quasars from the compilation of
\citet{schneider_sloan_2010} of all spectroscopically confirmed SDSS
quasars, that lie in the Northern Galactic Cap, that have an i-band
magnitude brighter than 20.4, and that have low Galactic extinction,
as determined by the maps of \citet{1998ApJ...500..525S}.
Avoiding the Southern Galactic Cap means that the sample does not
contain the various ``special'' samples of quasars targeted on the
Celestial Equator in the Fall sky [\citet{AdelmanMcCarthy:2005se}],
which tend to be more unusual, fainter, or less representative of the
quasar population as a whole.  The magnitude limit also removes those
objects at lowest signal-to-noise ratio. Indeed, the vast majority of
the $10^4$ objects are selected using the uniform criteria described
by \citet{2002AJ....123.2945R}.
The SEDs of these objects were
measured in the interval 3,793 {\AA} $< \lambda <$ 9,221 {\AA}, 
with a spectral resolution of $R \simeq$ 2,000 and accurate spectrophotometry
[\citet{adelman-mccarthy_sixth_2008}]. 
The number of quasars as a function of redshift in our sample is
shown in the left (red in color version) bars of Fig. \ref{Fig:SelFun},
and reproduces the redshift distribution of the 
SDSS quasar catalog as a whole rather well.

Starting from the spectra of our sample, we constructed synthetic 
fluxes using the 42 transmission functions
shown in Fig. \ref{Fig:filters}. The reduction is straightforward:
the flux is obtained by the convolution of the SDSS spectra
with the filter transmission functions:
$$
f_{a(p)} = \frac{1}{n_a} \int T_a(\lambda) S_p(\lambda) d\lambda \; ,
$$
where $f_{a(p)}$ is the flux of the object $p$ measured in the narrow-band filter 
$a$, $T_a$ is the transmission function of the filter $a$, 
$n_a = \int T_a(\lambda) d\lambda$ is the total transmission normalization,
and $S_p$ is the SED of the object.
The noise in each filter in obtained by adding the noise in each spectral bin 
in quadrature.
% to produce the errors: 
%$$
%\delta f_{a(p)}^2 = \frac{1}{n_a} \int T_a (\lambda) \delta S_p^2 (\lambda) d\lambda \; ,
%$$
%where $\delta f_{a(p)}$ is the flux error in the filter $a$, and $\delta S_p$ is the
%estimated error in the SED (per pixel.) 
%The procedure outlined above generates observations with unrealistic errors, 
%similar to those that would be made with our 
%set of narrow-band filters.
%We have not included spectrophotometric calibration errors in our analysis.

\subsection{Simulated sample of quasars}

The procedure outlined above generates fluxes with errors which are
totally unrelated to the errors we expect in a narrow-band filter survey.
The magnitude depths (and the signal-to-noise ratios) of the original SDSS sample
are characteristics of that instrument, and corresponds to objects with $i < 19.1$ 
for $z<3.0$, and $i < 20.2$ for $z>3$. However, we want to determine
the accuracy of photo-z methods for a narrow-band survey that
reaches $i \sim 23$. Hence, we need a sample which
includes, on average, much less luminous objects than the SDSS catalog does. It is easy 
to construct an approximately fair sample of faint objects from a fair sample of bright objects, 
as long as the SEDs of these objects do not depend strongly on their luminosities -- 
which seems to be the case for quasars [\citet{1977ApJ...214..679B}].

We have used our original sample of 10,000 SDSS quasars described
in the previous Section to construct a simulated sample of quasars. 
For each object in the original sample with a
magnitude $i$ we associate an object in the simulated sample of 
magnitude $i_s$, given by:
\begin{equation}
\label{Eq:ii}
i_s = 14 + 1.5 ( i - 14) \; .
\ee
Since the original sample had objects with magnitudes 
$i \sim 14-20.5$, the simulated sample has objects ranging from
$i_s \sim 14$ to $i_s \sim 23.5$. The distribution of quasars as a function
of their magnitudes, in the original and in the simulated samples,
are shown in Fig. \ref{Fig:Mags}.
Clearly, Eq. (\ref{Eq:ii}) still reproduces the selection criteria of the original SDSS sample,
which is evidenced by the step-like features of the histograms shown in Fig.
\ref{Fig:Mags}. However, in this Section we are not as concerned with the
number of quasars as a function of redshift and magnitude (which we believe
are well represented by the luminosity function that was employed in the
previous Section), but with the accuracy of the photometric redshifts and 
the fraction of catastrophic outliers -- i.e., the instances when the photometric 
redshifts deviate from the spectroscopic redshifts by more than a given threshold. 
While we have not detected any significant correlations between the absolute or relative
magnitudes and the accuracy of the photo-z's, we have found that the
number of photo-z outliers is higher for the simulated sample, compared with
the original sample, which means that the rate of outliers does depend to
some extent on the actual magnitudes of the sample. This is discussed in 
detail in Section \ref{Sec:Templates}.

\begin{figure}
\vspace{0.5cm}
\includegraphics[width=80mm]{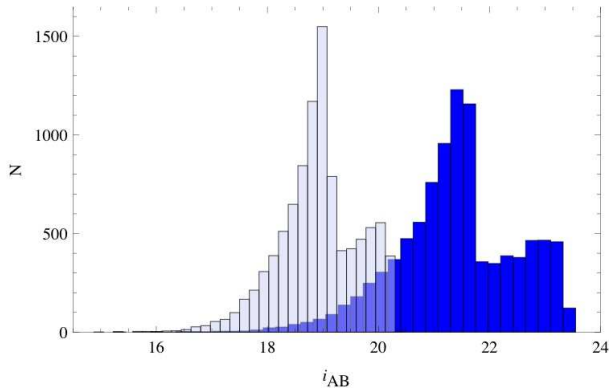}
\caption{
Distribution of magnitudes of the objects in our original sample
(light bars) and in the simulated sample (dark bars.)
}
\label{Fig:Mags}
\end{figure}

In order to generate realistic signal-to-noise ratios (SNR) for the objects
in this simulated sample, we also need to specify the depths of the survey
that we are considering, in each one of its 42 filters.
The 5$\sigma$ magnitude limits that we have estimated for J-PAS,
considering the size of the telescope, an aperture of 2 arcsec,
the median seeing at the site, the total exposure 
times for an 8,000 deg$^2$ survey over 4 years, the presumed read-out 
noise, filter throughputs, night sky luminosity, lunar cycle, etc., are shown 
in Fig. {\ref{Fig:depths}}.
\begin{figure}
\vspace{0.5cm}
\includegraphics[width=80mm]{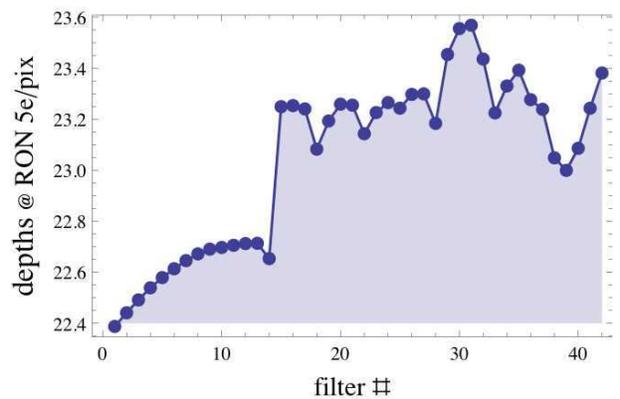}
\caption{
Estimated limiting magnitudes ($5\sigma$) for J-PAS with an aperture of 2'', 
assuming a read-out noise of 5e/pixel.
}
\label{Fig:depths}
\end{figure}

Our model for the signal-to-noise ratio (SNR) in each filter, for simulated
quasars of a given i-band magnitude $i_s$, is the following:
\be
\label{Eq:SNR}
SNR(a) = 5 \, \frac{f(a)}{\bar{f}_i} \, 10^{0.4[d(a)-i_s]} \; ,
\ee
where $\bar{f}_i$ is the average flux of that object in the 
10 filters (7,100 $\leq \lambda \leq$ 8,100) that overlap with the 
$i$-band; $f(a)$ is the flux in filter $a$;
$d(a)$ is the $5\sigma$ depth of filter $a$ from Fig. {\ref{Fig:depths}}; and 
$i_s$ is the (simulated) $i$-band magnitude of that object. 
This model assumes that the intrinsic photon counting noise of the
quasar is subdominant compared to other sources of noise such as the
sky or the host galaxy.
In order to obtain the desired SNR in our simulated sample,
we have added a white (Gaussian) noise to the fluxes of the original
sample, such that the final level of noise is the one prescribed by 
Eq. (\ref{Eq:SNR}).

\subsection{Photometric redshifts of quasars: Template Fitting Method}
\label{Sec:Templates}

Conceptually, fitting a series of photometric fluxes to a template is
the simplest method to obtain redshifts from objects that belong to 
a given spectral class [\citet{benitez_bayesian_2000}]. 
Type-I quasars possess a (double) power-law continuum that
rises rapidly in the blue, and a series of
broad ($\Delta \lambda/\lambda \sim$ 1/20 -- 1/10 FWHM) 
emission lines -- see Fig. \ref{Fig:quasars}. 
At high redshifts ($z \gtrsim 2.5$) the 
Ly-$\alpha$ break (which is a sharp drop in the observed
spectrum of distant quasars due to absorption from 
intervening neutral Hydrogen) can be seen at 
$\lambda \gtrsim$ 4,000 {\AA}, which lies just within the dynamic
range of the filter system we are exploring here. These
very distinct spectral features, which are
clearly resolved with our filter system, allow not only the
extraction of excellent photo-z's, but can also
be used to distinguish quasars from stars unambiguously 
-- see, e.g., the SDSS spectral templates, \citet{adelman-mccarthy_sdss_2008}. 
%% ADDED NEW
The COMBO-17 quasar catalog [\citet{Wolf_COMBO17_q}] has successfully
employed a template fitting method not only to obtain photometric redshifts,
but also to identify stars and understand the completeness and rate of
contamination of the quasar sample.

Here we will assume that all quasars have already been identified,
and the only parameter that we will fit in our tests is the redshift of a given object.
A more detailed analysis will be the subject of a forthcoming publication
(Gonzalez-Serrano {\it et al.}, 2012, to appear).

Our baseline model for the quasar spectra is
the Vanden Berk mean template [\citet{vanden_berk_composite_2001}], 
which also includes the uncertainties due to intrinsic variations. 
We allow for further variability in the quasar spectra 
by means of the global eigen-spectra computed by \citet{yip_spectral_2004}.
We use both the uncertainties in the Vanden Berk template and the
Yip {\it et al.} eigen-spectra because they capture different types of intrinsic
variability: while the uncertainties in the template are more suited to allow for
uncorrelated variations around the emission lines and below the 
Ly-$\alpha$, the Yip {\it et al.} eigen-spectra allow for features
such as contamination from the host galaxy (which is most relevant 
at low luminosities), UV-optical continuum variations, correlated Balmer 
emission lines and other secondary effects such as broad absorption 
line systems.
We search for the best-fit combination of the four eigen-spectra 
at each redshift, by varying their weights ($w_{p,z}$ , $p=1 \cdots 4$) 
in the interval $-3 w_{p} \leq w_{p,z} \leq 3 w_{p}$, where $w_p$ 
is the weight of the $p$-th eigenvalue relative to the mean. 
The four highest-ranked global eigen-spectra have weights of $ w_1 = 0.119$, 
$ w_2 = 0.076$, $ w_3 = 0.066$, and $ w_4 = 0.028$ relative to the
mean template spectrum (which has $w_0=1$ by definition) [\citet{yip_spectral_2004}].

The eigen-spectra are included in the MLE 
in the following way: first, we normalize the fluxes by their 
square-integral, i.e.: 
$f_{a} \rightarrow f^n_a = f_a/\sqrt{ \sum_{b=1}^{N} f_b^2}$,
where $N$ is the number of filters (42 for J-PAS.)
We then add the redshifted eigen-spectra $f^n_{p,a}(z)$ to
the average template [$f^n_{0,a}(z)$] with weights $w_{p}(z)$,
so that at each redshift we have  
$f_{a}^n = f^n_{0,a} + \sum_{p=1}^4 w_{p} f^n_{p,a}$. The
weights $w_p$ are found by minimizing the 
(reduced) $\chi^2$ at each redshift:
\be
\label{Eq:chi2}
\chi^2(i,z) = \frac{1}{N} \sum_a^{N} \frac{ 
\left[ f_{a}^n - f_{a}^n(i) \right]^2}{ \sigma_{a}^2(i,z) } \; ,
\ee
where $f_{a}^n(i)$ are the fluxes from some object $i$ in our sample
of SDSS quasars, and $\sigma_{a}^2(i,z)$ is the sum in quadrature of the 
flux errors and of the (2-$\sigma$) uncertainties in the quasar template 
spectrum for that filter.
We have not marginalized over the weights of the eigen-spectra -- 
i.e., the method is indifferent as to whether or not the best fit to an object 
at a given redshift includes an unusually large contribution 
from some particular eigen-mode.

It is also interesting to search for the linear combination between the 
fluxes that leads to the most accurate photo-z's.
We could have employed either the fluxes themselves 
or the colors (flux differences) for the procedure that was outlined above --
or, in fact, any linear combination of the fluxes. Most photo-z methods employ colors  
[\citet{benitez_bayesian_2000,blake_cosmology_2005,
Firth_2003, budavari_unified_2009}], since this seems to
reduce the influence of some systematic effects such as reddening,
and it also eliminates the need to marginalize over the normalization of the observed flux.
We have tested the performance of the template fitting method using the fluxes $f_a$,
the colors $\Delta f_{a} = f_a - f_{a-1}$ (the derivative of
the flux), and also the second differences $\Delta^2 f_{a} = f_{a+1} - 2 f_a + f_{a-1}$
(the second derivative of the flux, or color differences.)
We have noticed a slightly better performance with the latter choice ($\Delta^2 f_a$)
when compared with the usual colors ($\Delta f_a$), but the difference is negligible
and therefore in this work we have kept the usual practice of using colors.
The results shown in the remainder of this Section
refer to the traditional template-fitting method with colors.

In Fig. \ref{Fig:HistoChi2} we plot
the distribution of $\log_{10} \chi^2$ (for the best-fit 
$\chi^2$ among all $z$'s) for our sample of $10^4$ quasars.
The wide variation in the quality of the fit is partly due to the
small number of free parameters: we fit only the redshift and the 
weights of the four eigen-modes.
% Since real quasars inside real galaxies cannot be described only by 
% four degrees of freedom at all redshifts, the fits can be quite poor 
% for some objects.

\begin{figure*}
\vspace{0.5cm}
\includegraphics[width=75mm]{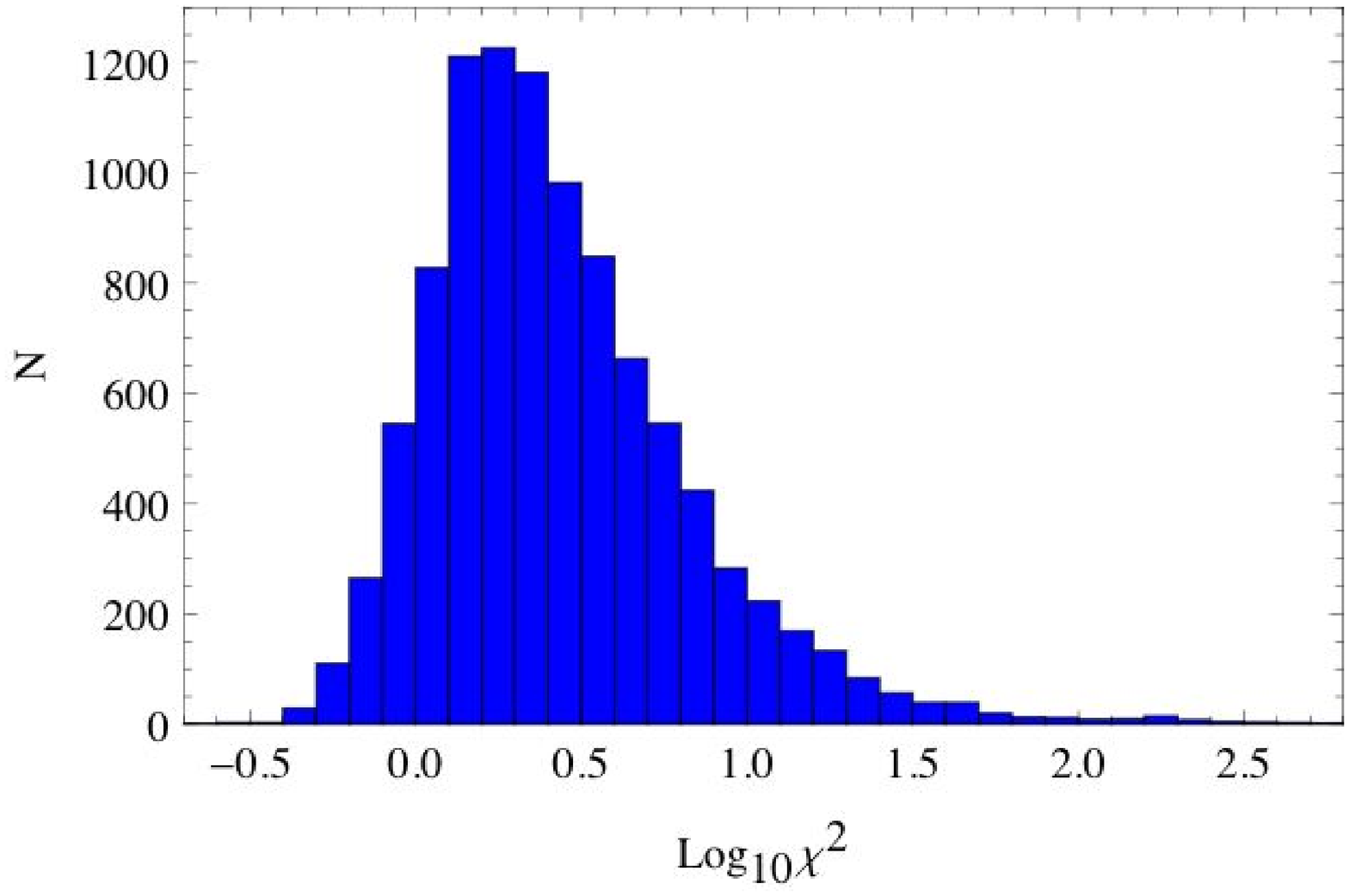}
\hspace{0.5cm}
\includegraphics[width=75mm]{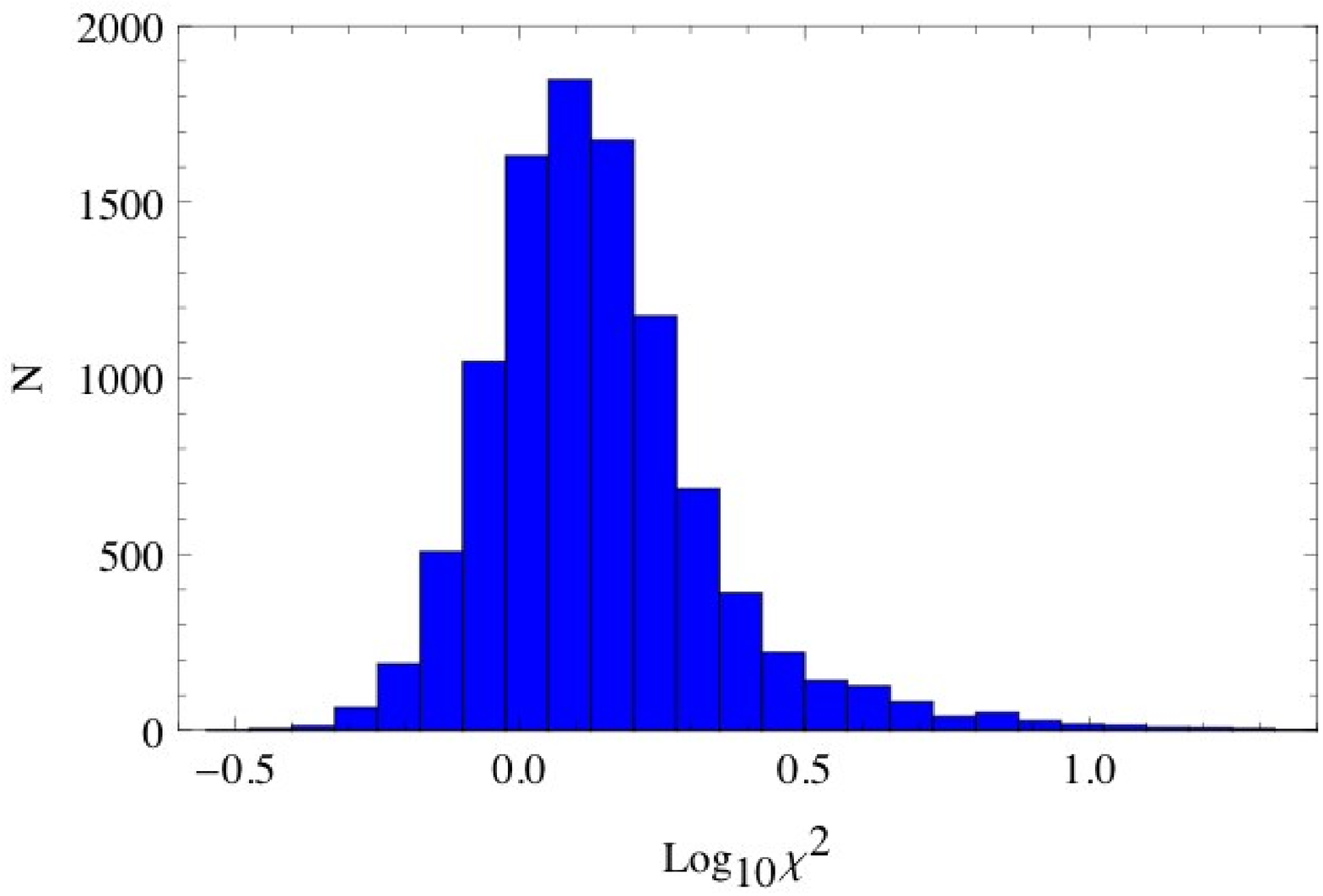}
\caption{
Histogram of the best-fit reduced $\chi^2$ for the sample of $10^4$ quasars from
the SDSS spectroscopic catalog. Left panel: original SDSS sample limited
at $ i \lesssim 20.1 $; right panel: simulated sample, effectively limited at $ i \lesssim 23.5 $.
%The simulated sample shows a narrower distribution because the more 
%peculiar objects fit the template more easily when the data have larger errors.
We point out that the distributions above are not at all typical of a $\chi^2$ probability
distribution function -- 
the horizontal axis is in fact $\log_{10} \chi^2$.
}
\label{Fig:HistoChi2}
\end{figure*}

Once the $\chi^2(z)$ has been determined for a given object,
we build the corresponding posterior probability distribution function (p.d.f.):
\be
\label{Eq:Prob}
p(z) \propto e^{-\chi^2(z)/2} \; .
\ee
The photometric redshift is the one that minimizes the $\chi^2$ 
(the MLE.)

Finally, we need to estimate the ``odds'' that the photo-z of a given
object is accurate. Due to the many possible mismatches between 
different combinations of the emission lines, the p.d.f.'s are highly 
non-Gaussian, with multiple peaks (i.e., multi-modality.)
Hence, we have employed an empirical set of indicators to 
assess the quality of the photo-z's.
These empirical indicators are: (i) the value of the best-fit $\chi^2$;
(ii) the ratio between the posterior p.d.f. $p(z)$ at the
first (global) maximum of the p.d.f. and 
the value of the p.d.f. at the secondary maximum (if it exists), 
$r= p_{max \# 1}/p_{max \# 2}$; and
(iii) the dispersion of the p.d.f. around the best fit, $\sigma = \int (z-z_{best})^2 p(z) dz$.
We then maximize the correlation between the redshift
error $ |z_p-z_s|/(1+z_s)$ and a linear combination of 
simple functions of these indicators.
Finally, we normalize the results so that they lie between $0$
(a very bad fit) and $1$ (very good fit.)

For the original SDSS sample, we found empirically that
the combination that correlates (positively)
most strongly with the photo-z errors (the quality) is given by:
\be
{\rm q} = 0.15 \log (0.7 + \chi^2_{bf}) +  
e^{8 (r-1)} + 0.06 \, e^{1.4 \sigma} \; .
\label{Eq:Odds1}
\ee
For the simulated sample, the quality indicator is:
\be
{\rm q} = 0.3 \log (0.6 + \chi^2_{bf}) +  e^{15 ( r-1)} + 0.026 \, e^{\sigma} \; .
\label{Eq:Odds2}
\ee

Finally, we compute the quality factor $0 < \bar{q} \leq 1$ with the formula:
\be
\bar{q} = \left[ \frac{{\rm max}(q)-q}{{\rm max}(q)- {\rm min}(q)} \right]^4 \; ,
\ee
where the power of $4$ was introduced to produce a ``flatter''
distribution of bad and good fits (this step does not
affect the photo-z quality cuts that we impose below).
%Notice that this formula ensures that the quality $\bar{q}$ lies strictly between
%0 and 1. These empirical formulas should be recalibrated in light of new 
%data and further refinements of the method.

The relationship between the quality factor and the photometric
redshift errors is shown in the distributions of Fig. \ref{Fig:histo}. 
There is a strong correlation between the quality factor and the rate of
``catastrophic errors'', which we define arbitrarily
as any instance in which $ |z_{p}-z_{s}|/(1+z_{s}) \geq 0.02$
-- denoted as the horizontal dashed lines in Fig. \ref{Fig:histo}.
We have adopted the usual convention of scaling the redshift errors
by $1+z$, since this is the scaling of the rest-frame spectral features. 
There is no obvious reason why emission-line systems (whose salient 
features can enter or exit the filter system depending on the redshift) 
should also be subject to this scaling, but we have verified that the 
scatter in the non-catastrophic photo-z estimates do indeed scale 
approximately as $1+z$.

\begin{figure*}
\vspace{0.5cm}
\includegraphics[width=75mm]{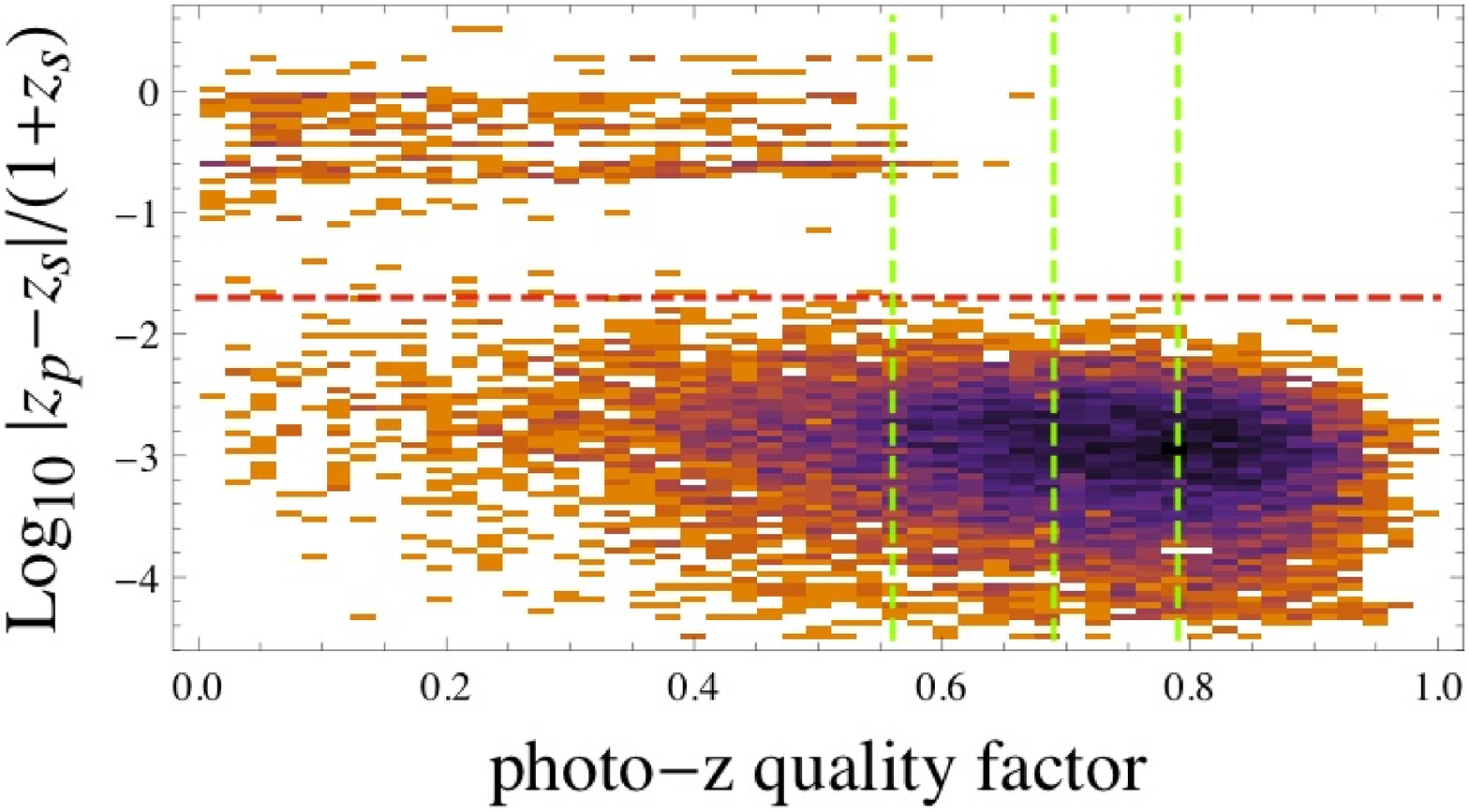}
\hspace{0.5cm}
\includegraphics[width=75mm]{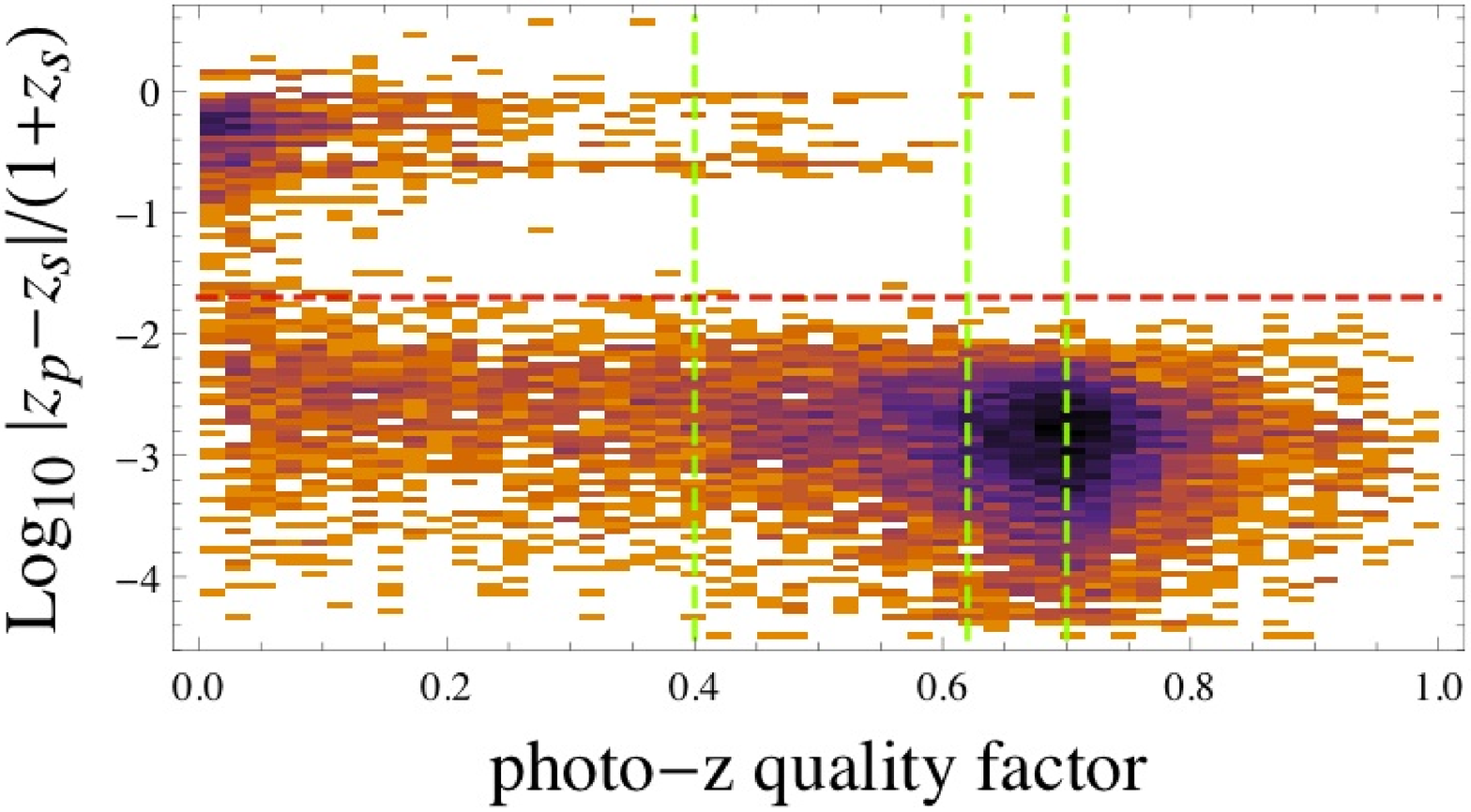}
\caption{2D histograms of the photo-z errors
$\log_{10} |z_{p}-z_{s}|/(1+z_{s})$ (vertical axis)
and the quality factor $\bar{q}$ (horizontal axis). The left and right panels
correspond to the original and the simulated samples, respectively. 
The catastrophic redshift errors 
[$|z_{p} - s_{s}|/(1+z_{s}) \geq 0.02$] lie above the 
horizontal dashed (red in color version) line.
The quality factor has been grouped into four ``grades'',
from grade=1 to grade=4, according to the vertical dashed 
(green in color version) lines.
}
\label{Fig:histo}
\end{figure*}

We have divided our sample into four groups with an equal number of objects, 
according to the value of $\bar{q}$: 
lowest quality ($g_1$, 2,500 objects), medium-low quality
($g_2$, 2,500 objects),
medium-high quality ($g_3$, 2,500 objects) and highest quality ($g_4$, 2500 objects) 
photo-z's. 
These grade groups are separated by the vertical dotted lines shown in 
Fig. \ref{Fig:histo}. 
For the original sample, the rate of catastrophic redshifts 
is 16.9 \%, 0.08 \%, 0 \% and 0 \% in the grade groups $g_1$, 
$g_2$, $g_3$ and $g_4$, respectively. 
For the simulated sample, the rate of catastrophic errors 
is 44.7 \%, 2.3 \%, 0.001 \% and 0 \% in the groups $g_1$, 
$g_2$, $g_3$ and $g_4$, respectively.

The relationship between spectroscopic and photometric redshifts
is shown in Fig. \ref{Fig:scatterplot}, 
where each quadrant corresponds to a grade group.
Almost all the catastrophic redshift errors are in the $g_1$ grade
group, and most of the catastrophic errors lie
below $z_p \lesssim 2.5$ -- since it is above this redshift that the 
Ly-$\alpha$ break becomes visible in our filter system.
% The bad photo-z's are also slightly biased towards low 
% redshifts, mainly because it is harder to obtain a low
% $\chi^2$ for high-redshift quasars due to their more prominent spectral features
% (in particular the Ly-$\alpha$ break.)

From Figs. \ref{Fig:histo} and \ref{Fig:scatterplot} it is clear that the rate of
catastrophic photo-z's is larger for the simulated sample, which has an overall
fraction of approximately 12\% of outliers, compared to the original sample, which has a
total fraction of 4\% of outliers. A similar increase happens
also when the Training Set method is applied to these samples (see the next Section).
Since the simulated sample used in this Section was not designed to reproduce the 
actual distribution of magnitudes expected in a real catalog of quasars, this 
means that our results for the rate of outliers are only an estimate for the
actual rate that we should expect from the final J-PAS catalog. 
However, even as the rate of outliers increases from the original to the simulated 
samples, the accuracy of the photo-z's are still very nearly the same.
This means that the actual distribution of magnitudes of an eventual J-PAS quasar 
catalog should have little impact on the accuracy of the photo-z's -- although it could
affect the completeness and purity of that catalog.

A further peculiarity of the quasar photo-z's is 
evident in the lines $z_{p} = z_* + \alpha z_{s}$, 
which are most prominent in the $g_1$ groups of
the original and simulated samples, as well as the $g_2$ group
of the simulated sample.
%These source of these ``degeneracy lines" is in fact related to 
%our choice of the maximum likelihood rather than an expectation 
%value $\langle z \rangle = \int dz \, z \, p(z)$. Our preference for the
%maximum likelihood method was motivated by the observation
%that, for quasars, very frequently the p.d.f. $p(z)$ is bi- or even 
%multi-modal. This multi-modality is explained primarily by the nature of the 
%signature features of the average QSO spectrum, i.e.,
%the set of broad emission lines. 
Whenever two (or more) pairs of broad emisson lines are 
separated by the same relative interval in wavelength, i.e.
$\lambda_\alpha/\lambda_\beta \simeq 
\lambda_\gamma/\lambda_\delta$, 
(where $\lambda_{\alpha \cdots \delta}$ are the central
wavelengths of the emission lines), there is an enhanced
potential for a degeneracy of the fir between the data and 
the template -- i.e., additional peaks appear in the p.d.f. $p(z)$. 
As the true redshift of the quasar change, 
the ratios between these lines remain invariant, and 
so the ratios between the true and the false redshifts, 
$(1+z_{true})/(1+z_{false})$, also remain constant, giving rise to 
the lines seen in Fig. \ref{Fig:scatterplot}. The degeneracy is broken
when additional emission lines come into the filter system, which explains
why some redshifts are more susceptible to this problem.

\begin{figure*}
%\vspace{0.5cm}
\includegraphics[width=75mm]{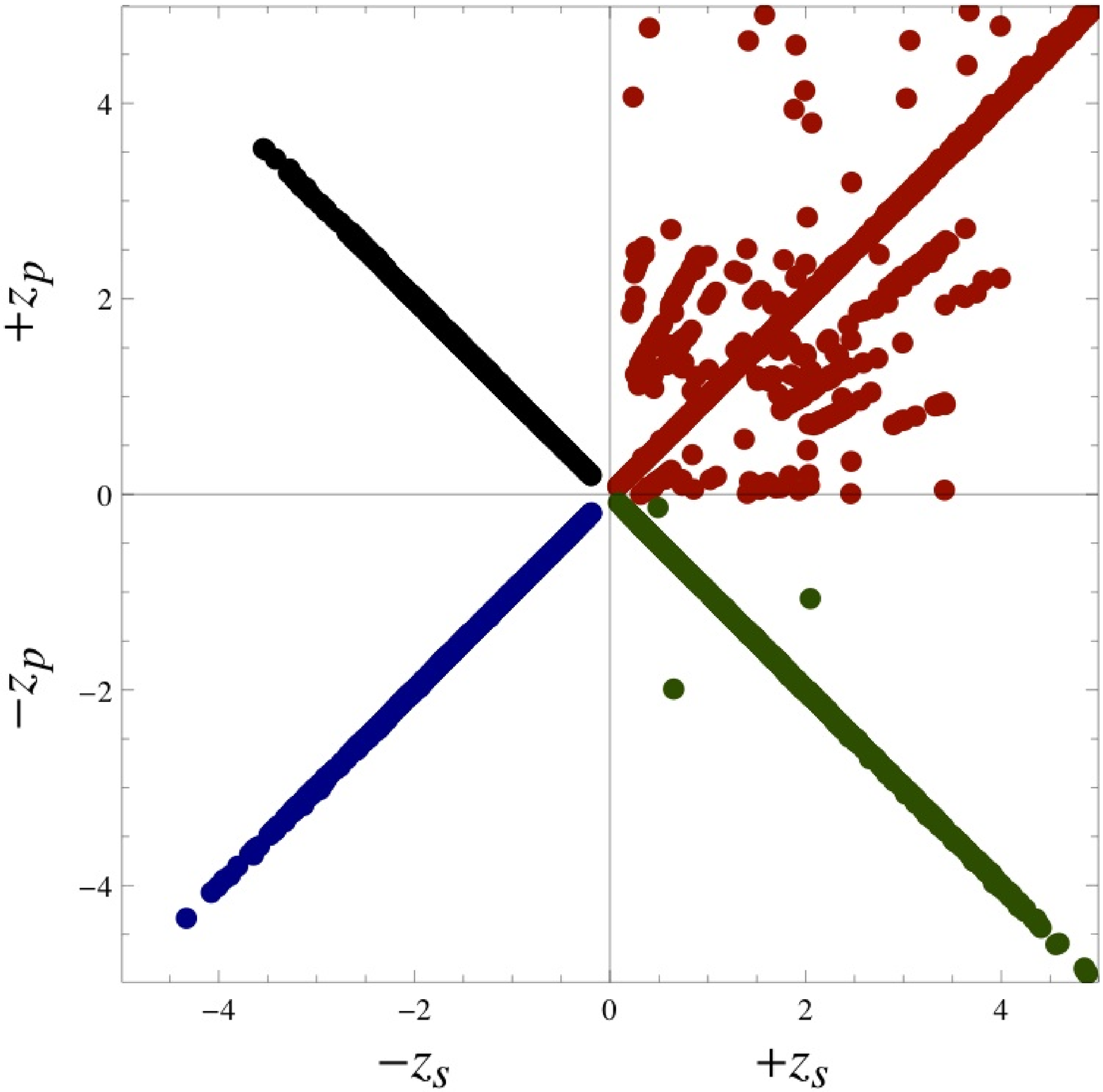}
\hspace{0.5cm}
\includegraphics[width=75mm]{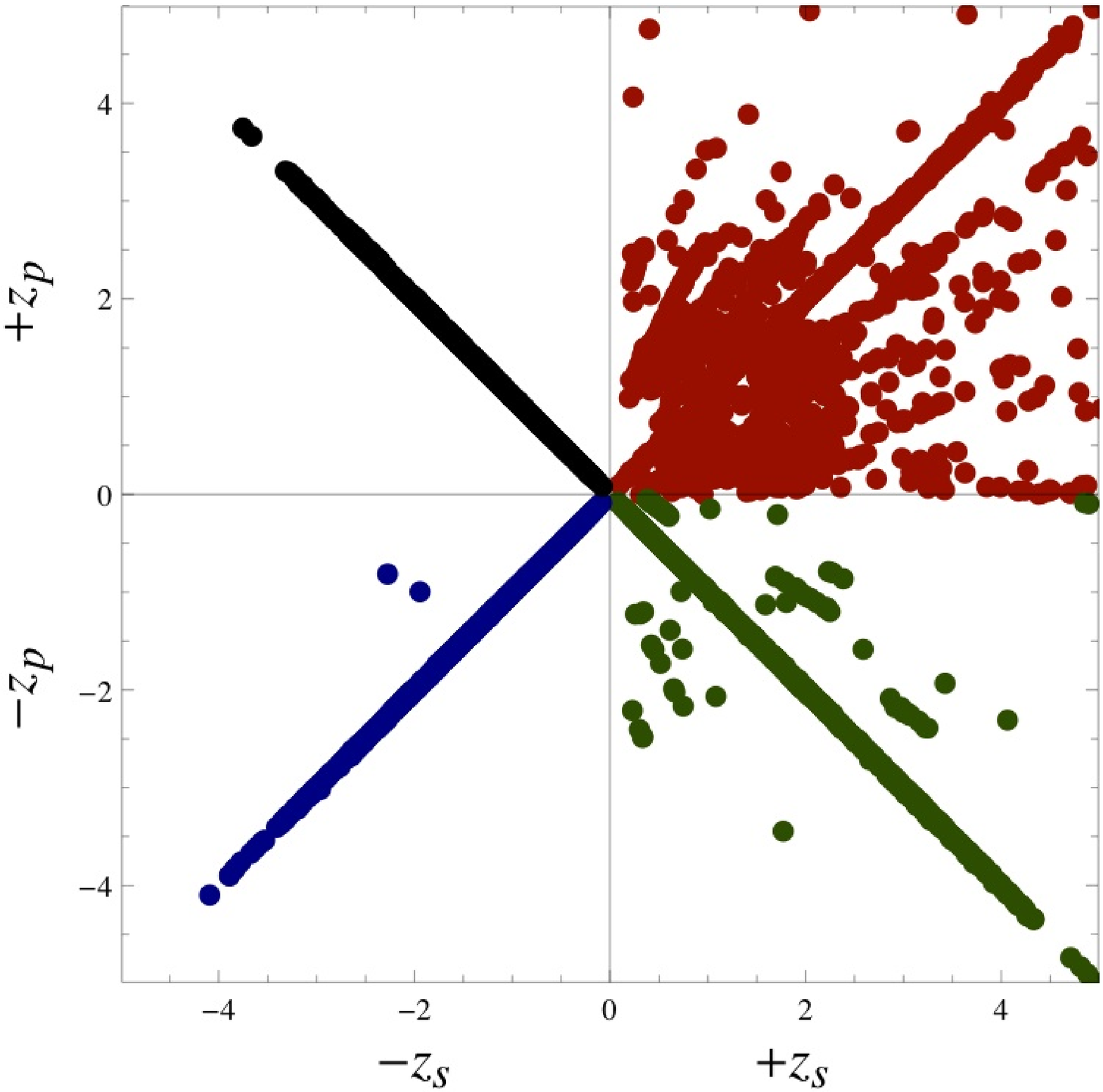}
\caption{Scatter-plots of spectroscopic redshifts (horizontal axis)
{\it versus} photometric redshifts (vertical axis) obtained with the template fitting method,
for the four quality grade groups (1, 2, 3 and 4). Left panel: original sample; 
right panel: simulated sample. There are 2,500 objects in the group $g_1$ 
(first quadrant in the upper right corner, red dots in color version); 
2,500 objects in the group $g_2$ (second quadrant and green dots); 
2,500 objects in the group $g_3$ (third quadrant and blue dots); and 2,500 objects in the
group $g_4$ (fourth quadrant and black dots).
The radial lines in the $g_1$ group
correspond to degenerate regions of the $z_{p}-z_{s}$ mapping.
There are virtually no catastrophic errors for $z_{p} \gtrsim 2.5$ objects
in the $g_2$, $g_3$ and $g_4$ grades in the simulated samples.
}
\label{Fig:scatterplot}
\end{figure*}

The median and median absolute deviation ({\it mad}) of the redshift errors in each
grade groups are shown in Fig. \ref{Fig:medians}, for the original (left panel) and simulated
(right panel) samples. For the lowest quality photo-z's (grade group 1), the median for
the original sample of quasars is $med [ |z_p - z_s|/(1+z_s) ]  = 0.0019$, and the deviation is 
$mad [ |z_p - z_s|/(1+z_s) ]  = 0.0014$, which is very small given the high level of 
contamination from outliers -- 12\% for that group. For the simulated sample the redshift 
errors are much larger: the median and median deviation for group 1 are 
0.0073 and 0.0069, respectively -- which is not surprising given that the number
of catastrophic photo-z's is 44.7\%.
However, for the grade group 2 the median and median deviation for the original sample
falls to 0.001 and 0.0007, respectively. 
More importantly, for the simulated sample the median and deviation are
0.0014 and 0.001, respectively.
The accuracies of the photo-z's for the grade groups 3 and 4 are slightly higher still.

\begin{figure*}
%\vspace{0.5cm}
\includegraphics[width=75mm]{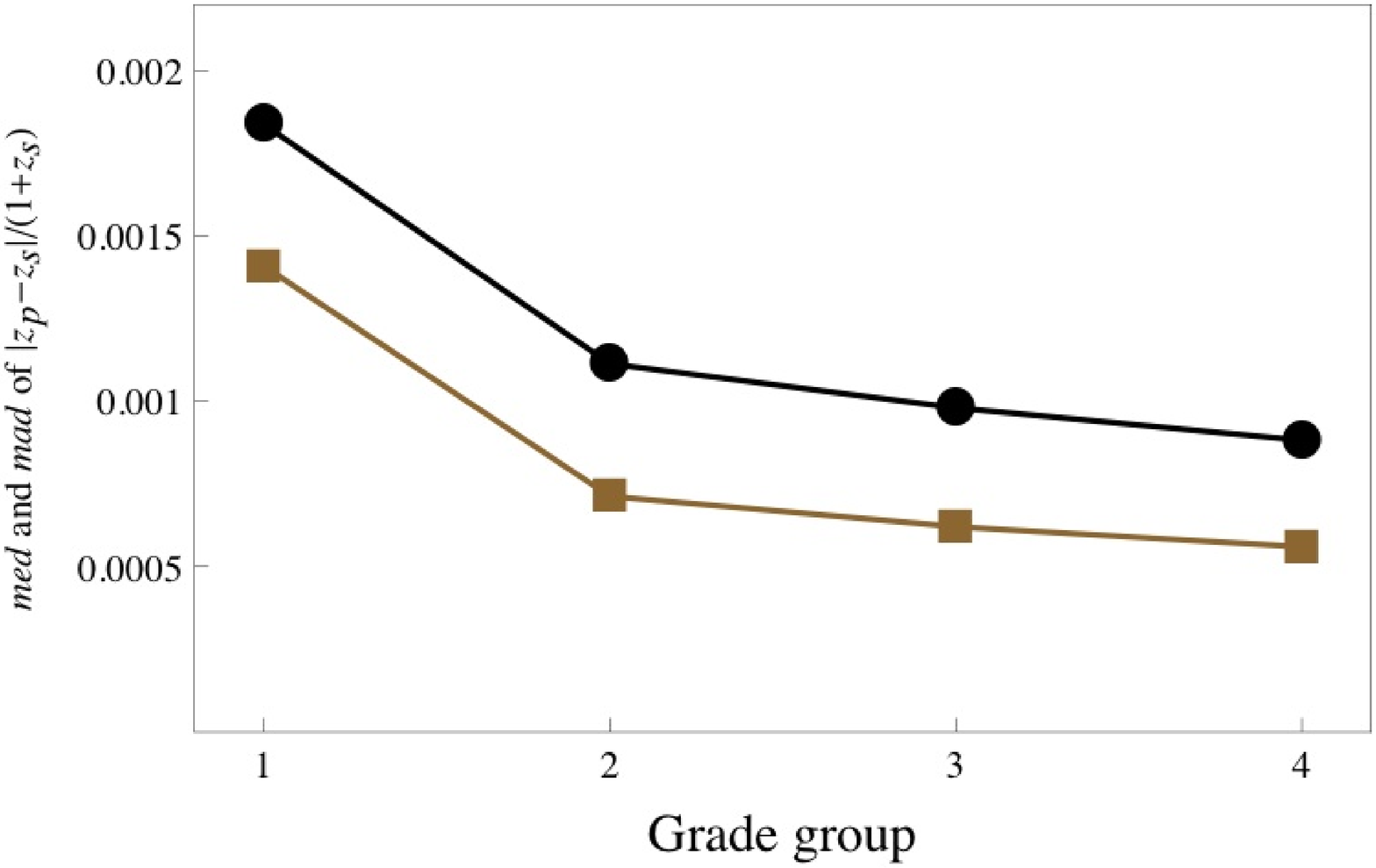}
\hspace{0.5cm}
\includegraphics[width=75mm]{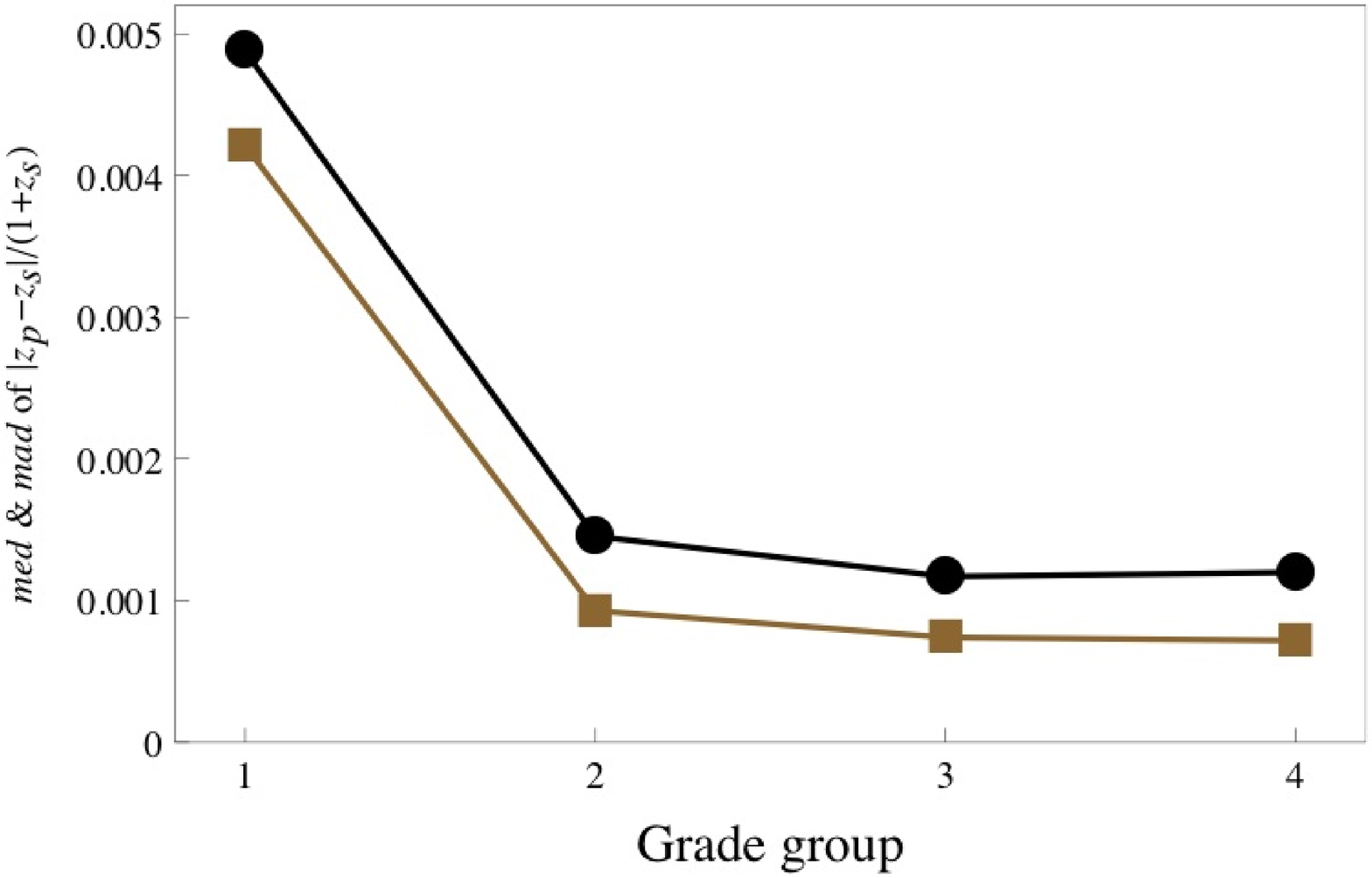}
\caption{Median ({\it med}) and median absolute deviation ({\it mad}) of the
errors in the photometric redshifts obtained with the template fitting method. 
Left panel: original sample of SDSS quasars; 
right panel: simulated sample. The circles (black in color version) denote
the medians for each grade group; squares (brown in color version) denote
the {\it mad}.
}
\label{Fig:medians}
\end{figure*}

An alternative metric to assess the accuracy of the photometric redshifts 
is to manage the sensitivity to catastrophic outliers with the following method.
First, we compute the {\it tapered}  (or bounded) error estimator defined by:
\begin{eqnarray}
\label{Eq:TapErr}
\left(\frac{\sigma^{T}_z}{1+z} \right)^2  =
\left\langle \left[ \delta_z \tanh \frac{1}{\delta_z}
\frac{z_{p} - z_{s}}{1+ z_{s}}
\right]^2 \right\rangle_{all} =
\\ \nonumber
\frac{1}{N}
\sum_i
\left[ \delta_z \tanh \frac{1}{\delta_z}
\frac{z_{p}(i) - z_{s}(i)}{1+ z_{s}(i)}
\right]^2 \; ,
\end{eqnarray}
where $ \delta_z = 0.02 $ in our case.
For accurate quasar photo-z's ($z_{p} \approx z_{s}$) with minimal
contamination from outliers, 
this error estimator yields the usual contribution to the rms error, while for 
samples heavily influenced by catastrophic photo-z's, this estimator 
assigns a contribution which asymptotes to our threshold $\delta_z.$

Second, we compute the purged rms error, summing only over
the non-catastrophic photo-z's:
\begin{eqnarray}
\label{Eq:NCErr}
\left( \frac{\sigma_z^{nc}}{1+z} \right)^2 = 
% \\ \nonumber
\frac{1}{N^{nc}}
\sum_{i=1}^{N^{nc}}
\frac{[z_{p}(i)-z_{s}(i)]^2}{[1+z_{s}(i)]^2} \; .
\end{eqnarray}
The estimators (\ref{Eq:TapErr})-(\ref{Eq:NCErr}) are therefore complementary: 
the tapered error estimator is indicative of
the rate of catastrophic errors, while the purged rms error is
a more faithful representation of the overall accuracy of the 
method for the bulk of the objects.
The results for the two estimators of the photometric
redshift uncertainties are shown in Fig. \ref{Fig:sigmas}, 
for the four grade groups. The two estimators are in good agreement
for the groups $g_2$, $g_3$ and $g_4$, which is again evidence
that the rate of catastrophic photo-z's is negligible for these groups.

Thus, we conclude that with the template fitting method alone it is possible to 
reach a photo-z accuracy better than 
$|z_s-z_p|/(1+z_s) \sim 0.0015$ for at least $\sim 75\%$ of quasars, 
even for a population of faint objects (our simulated sample), with a very small 
rate of catastrophic redshift errors. In fact, the average accuracy given by
the median and median deviation errors is already of the order
of the intrinsic error in the spectroscopic redshifts due to
line shifts [\citet{shen_clustering_2007,shen_catalog_2010}]. 
This means that, with filters of width $\sim$ 100 {\AA} (or, equivalently, 
with low-resolution spectroscopy with $R \sim 50$) we are saturating the
accuracy with which redshifts of quasars can be reliably estimated
-- although, naturally, with better resolution spectra and larger signal-to-noise
the rate of catastrophic errors would be even smaller.

%% ADDED - NEW
It is useful to compare the results of this section with those of the
COMBO-17 quasar sample [\citet{Wolf_COMBO17_q}].
That catalog, which employs five broad filters ($u g r i z$) and
12 narrow-band filters, attains a photo-z accuracy of $\sigma_z = 0.03$
-- the same that was also obtained for the COMBO-17 galaxy
catalog [\citet{Wolf_COMBO17_g}].
The accuracy that we obtain for quasars with the 42 contiguous 
narrow-band filters is also of the same order as that which is obtained
for red and emission-line galaxies [\citet{Benitez:2008fs}].
Clearly, the gains in photo-z accuracy are not linear with the width
of the filters, and the issue of continuous coverage over the entire 
dynamic range also plays in important role.

\begin{figure*}
%\vspace{0.5cm}
\includegraphics[width=75mm]{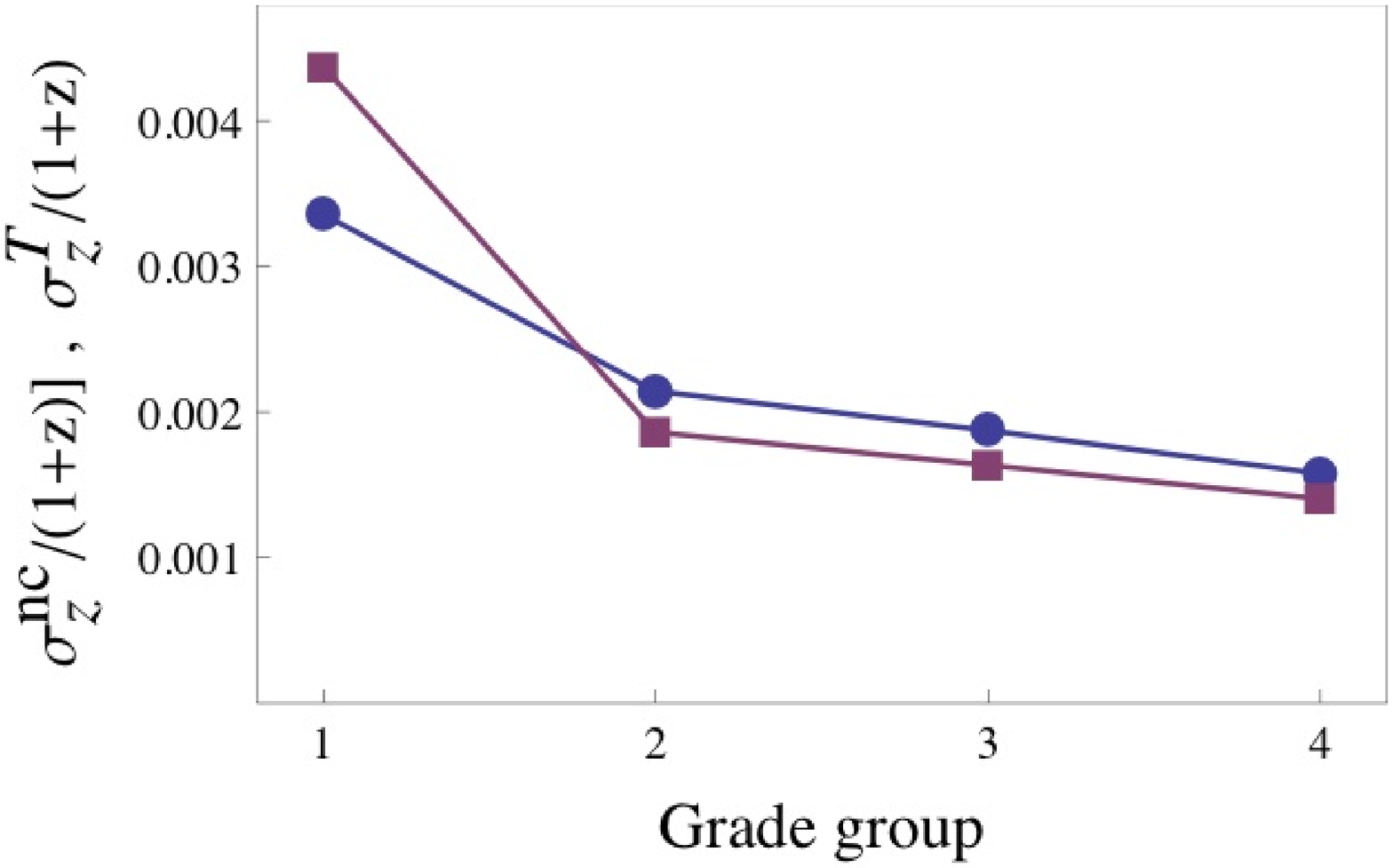}
\hspace{0.5cm}
\includegraphics[width=75mm]{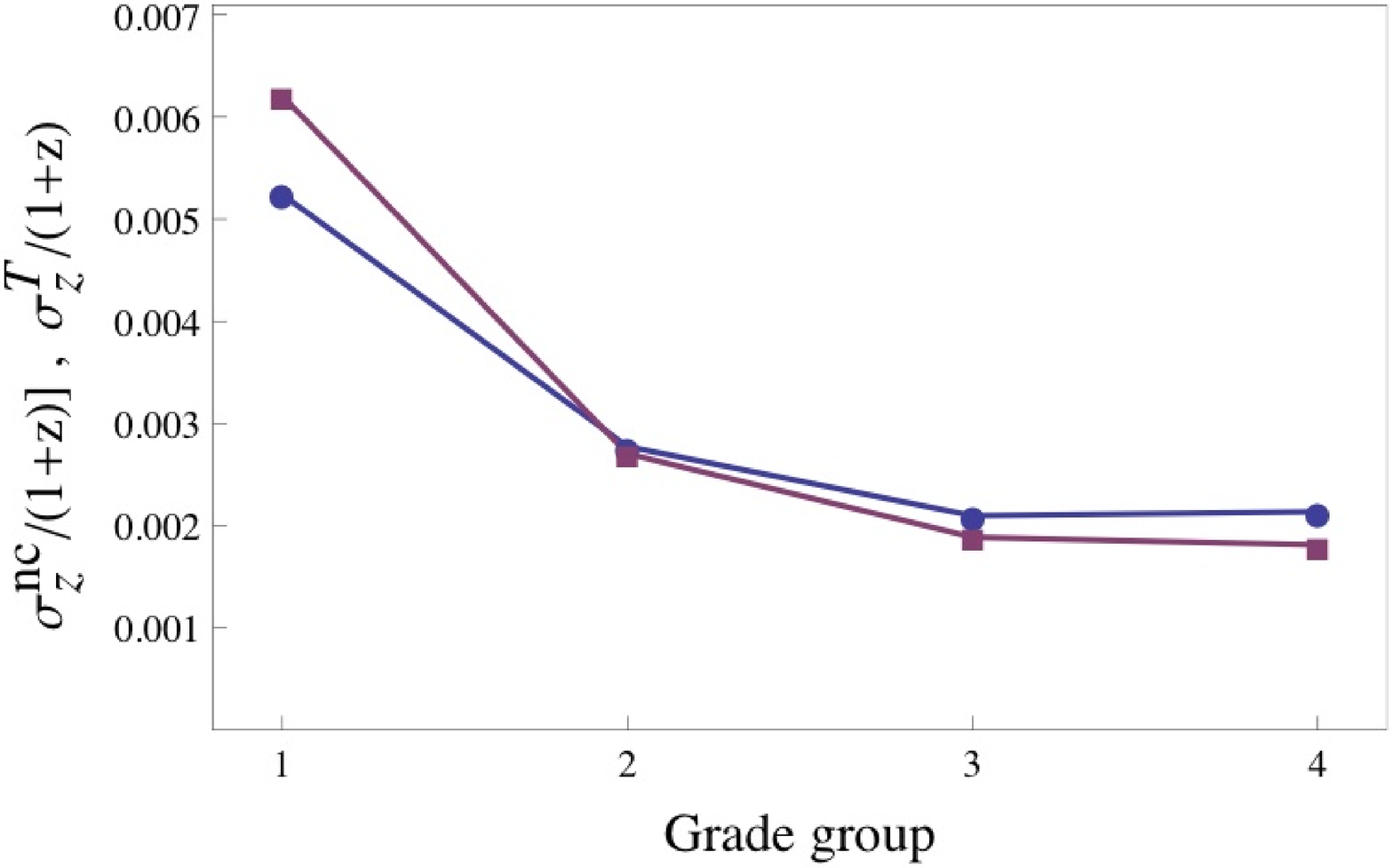}
\caption{Photo-z errors obtained with the template fitting method for each grade group: 
(i) circles (blue in color version): rms error excluding catastrophic redshift errors,
cf. Eq. \ref{Eq:NCErr};
and (ii) squares (red in color version): rms tapered error including
catastrophic redshift errors, cf. Eq. \ref{Eq:TapErr}.
When these two quantities coincide, 
the fraction of catastrophic photo-z's 
has become negligible.
}
\label{Fig:sigmas}
\end{figure*}

Finally, in order to understand how the photometric depth relates to photo-z depth,
it is useful to compare the photo-z quality indicator for each object to the i-band 
magnitude of the simulated sample, $i_s$, as well as the dependence of the actual 
photo-z errors with $i_s$. The magnitude is directly related to the SNR through Eq. (\ref{Eq:SNR}).
From the left panel of Fig. \ref{Fig:mags} 
(which should also be compared to the right panel of Fig. \ref{Fig:histo})
we see that the quality indicator declines steeply for the faintest objects
in the simulated sample. From the right panel of Fig. \ref{Fig:mags} 
we see that the actual photo-z errors (which are plotted on an inverted scale) 
also depend on the magnitude, but in this case even for the faintest 
objects a substantial fraction of the quasars still have correctly estimated redshifts.
This means that our quality indicator (which was calibrated for the full sample,
independently of magnitude) is not very good at capturing the photo-z dependence 
for the faintest objects.
Clearly, a more accurate analysis than the one we have implemented
can be achieved by including the magnitudes as additional parameters for estimating the 
photo-z's.

\begin{figure*}
\vspace{0.5cm}
\includegraphics[width=75mm]{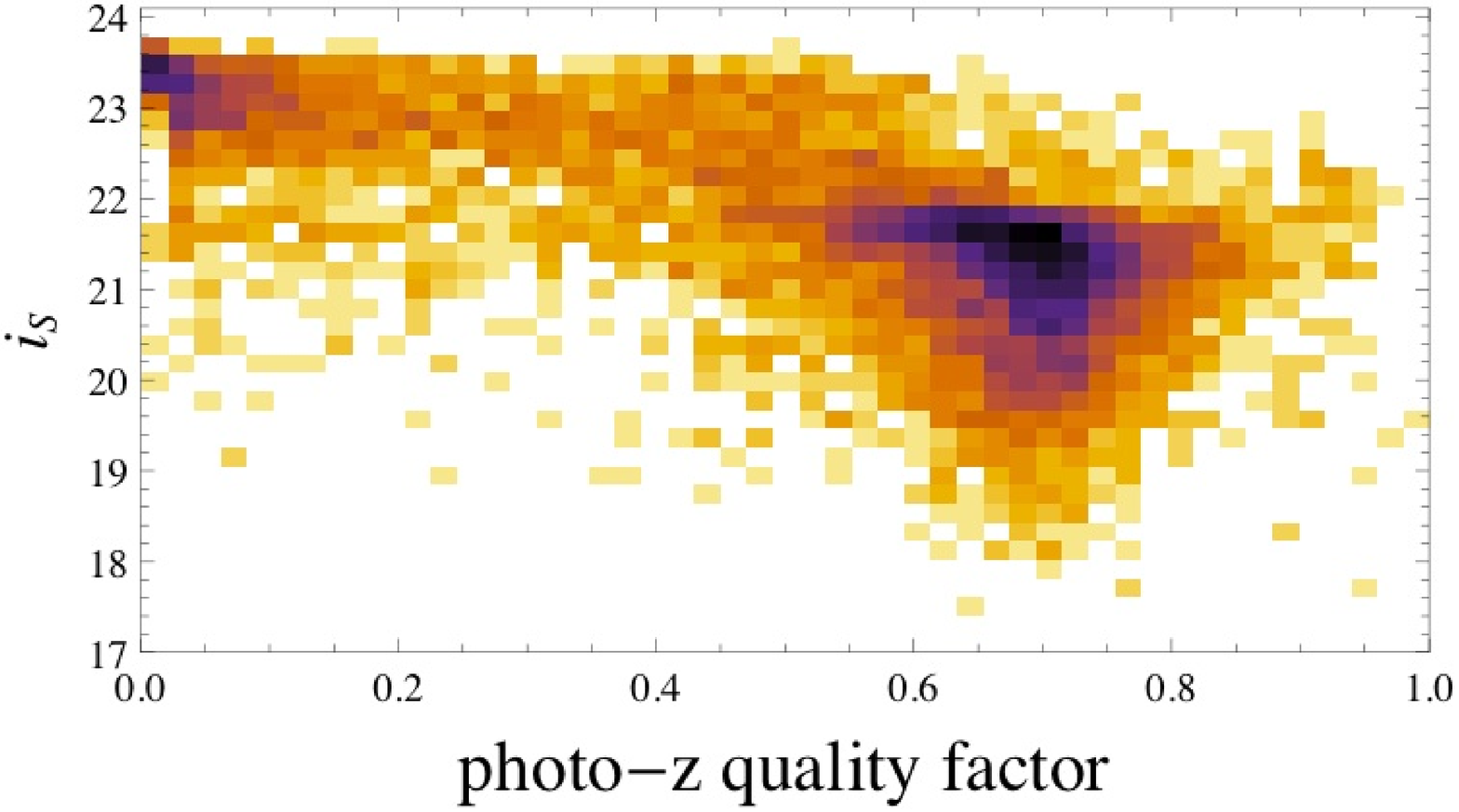}
\hspace{0.5cm}
\includegraphics[width=75mm]{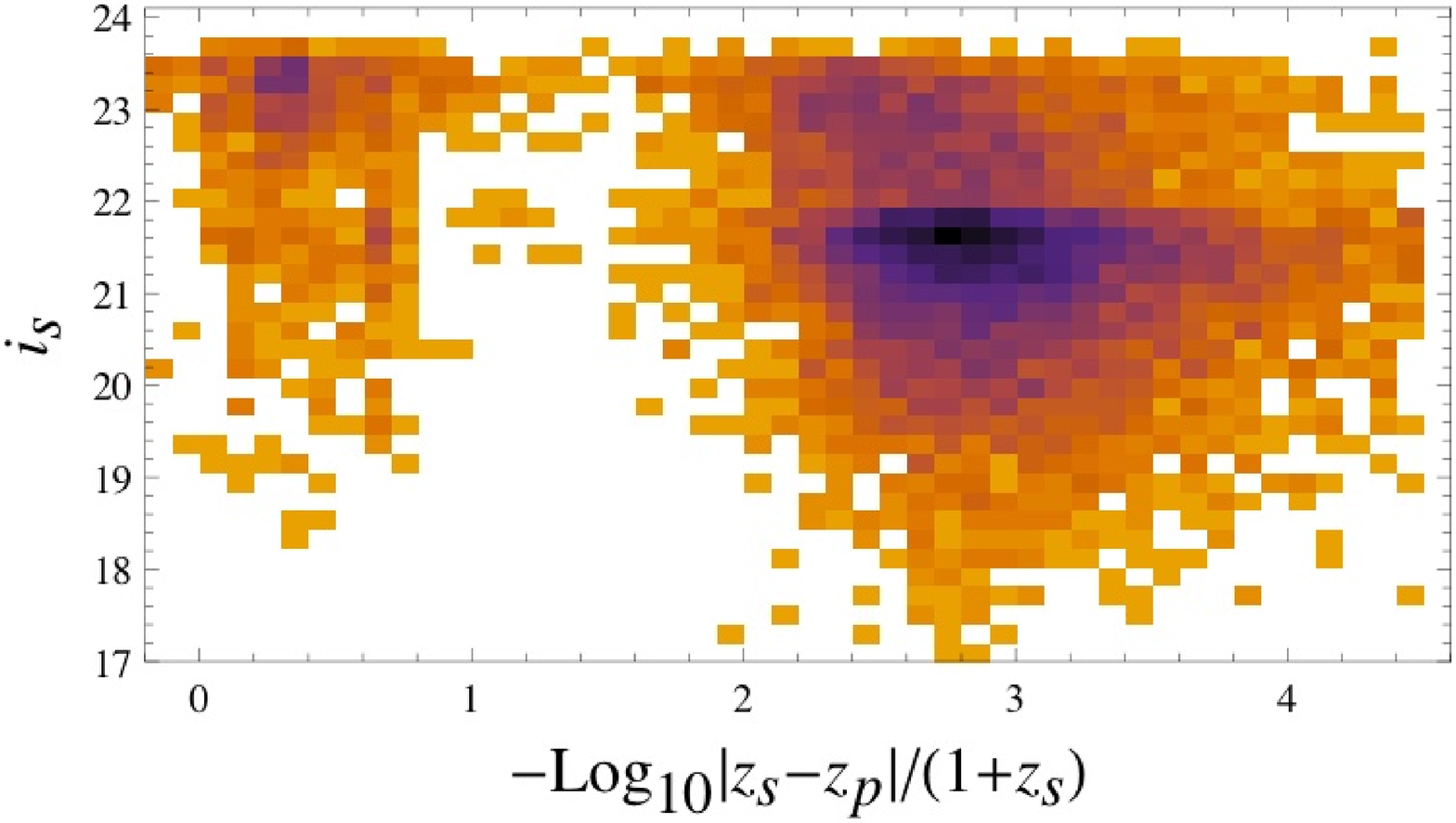}
\caption{2D histogram of the simulated sample, showing the magnitude in the $i$-band {\it versus} 
the photo-z quality indicator (left panel), and the magnitude {\it versus} the photo-z error on an 
inverted scale (right panel). These plots should be compared with the right panel of Fig. \ref{Fig:histo}.
}
\label{Fig:mags}
\end{figure*}

\subsection{Photometric redshifts of quasars: Training Set Method}

Training methods of redshift estimation are particularly well suited 
when a large and representative set of objects with known spectroscopic 
redshifts is available 
[\citet{Connolly_1995,Firth_2003,Csabai_2003,Collister_2004,
Oyaizu_2008,Banerji_2008,Bonfield_2010,Hildebrandt_2010}]. Ideally this training set must be a fair sample of 
the photometric set of galaxies for which we want to estimate redshifts, 
reproducing its color and magnitude distributions. Whereas lack of coverage in certain 
regions of parameter space may imply significant degradation in photo-z 
quality, having a representative and dense training set can lead to a superior 
photo-z accuracy compared to template fits. 
%, since the method naturally 
%captures the properties of the training set 
%and empirically reproduces them in the photometric sample.

Empirical methods use the training set objects to determine a
functional relationship between photometric observables (e.g. colors, magnitudes, types, 
etc.) and redshift. Once this function is calibrated, usually requiring that it reproduces 
the redshifts of the training set as well as possible, it can be straightforwardly 
applied to any photometric sample of interest. This class of methods includes machine 
learning techniques such as nearest neighbors [\citet{Csabai_2003}], local polynomial fits 
[\citet{Connolly_1995, Csabai_2003, Oyaizu_2008}], global neural networks 
[\citet{Firth_2003, Collister_2004, Oyaizu_2008}], and gaussian processes 
[\citet{Bonfield_2010}].
They have also been successfully applied to 
galaxy surveys, e.g. the SDSS [\citet{Oyaizu_2008}], allowing further  
applications in cluster detection [\citet{Dong_2008}] and weak lensing 
[\citet{Mandelbaum_2008, Sheldon_2009}]. 

The training set can also be used to improve template fitting, 
using it either to generate good priors or for empirical calibration and/or determination of the 
templates by, e.g., PCA of the spectra. Training sets are usually necessary to 
assess the photo-z quality 
of a certain survey specification and for calibration of the photo-z errors, which can then 
be modeled and included in a cosmological analysis [\citet{Ma_2006, Lima_2007}]. 
In this sense, it is the knowledge 
of the photo-z error parameters -- and not the value of the errors  themselves -- that limit 
the extraction of cosmological information from large data sets.

Here we implement a very simple empirical method, mainly to 
compare it with the template method presented in the previous Section.  We use 
a simple nearest neighbor (NN) method: for each photometric quasar, we 
search the training set for its nearest neighbor in magnitude space, and then
assign that neighbor's spectroscopic redshift as the best estimate for the photo-z 
of the photometric quasar. We define distances with an Euclidean 
metric in multidimensional magnitude space, such that the distance $d_{ij}$ between 
objects $i$ and $j$ is:
\be
d_{ij}^2=\sum_{a=1}^{N} (m^a_i-m^a_j)^2 \; ,
\ee
where $N=42$ is the number of narrow filters and $m^a_i $ is the $a^{\rm th}$ magnitude of 
the $i^{\rm th}$ object. 
The nearest neighbor to a certain object $i$ is then simply the object $j$ for which $d_{ij}$ is 
minimum.

We computed photo-zs in this way for all $10^4$ quasars in the catalog. For 
each quasar, we took all others as the training set. In this case, 
there is no need to divide the objects into a training and photometric set, 
because all that matters is the nearest neighbor. 
% Making such a division 
% would only decrease the training set density even further and prevent us 
% from evaluating photo-zs for all objects in the sample. Computing photo-zs for 
% all objects allows us to compare this method to the template fit photo-zs.

We can also use knowledge of the distance between the nearest neighbor and the 
second-nearest neighbor to assign a quality to the photo-z's obtained with the training
set method. The idea is that the quality of the photo-z is related to how sparse the 
training set is in the region around any given object.
The original and simulated samples
were then divided into four groups of increasing density (i.e., decreasing sparseness), 
as we did for the template fitting method.
In Fig. \ref{Fig:scatterplot2} we show the photo-z's as a function of spectroscopic 
redshifts for the original sample of quasars (left panel), and for the sample simulated
with J-PAS specifications (right panel), for the four quality groups.

\begin{figure*}
%\vspace{0.5cm}
\includegraphics[width=75mm]{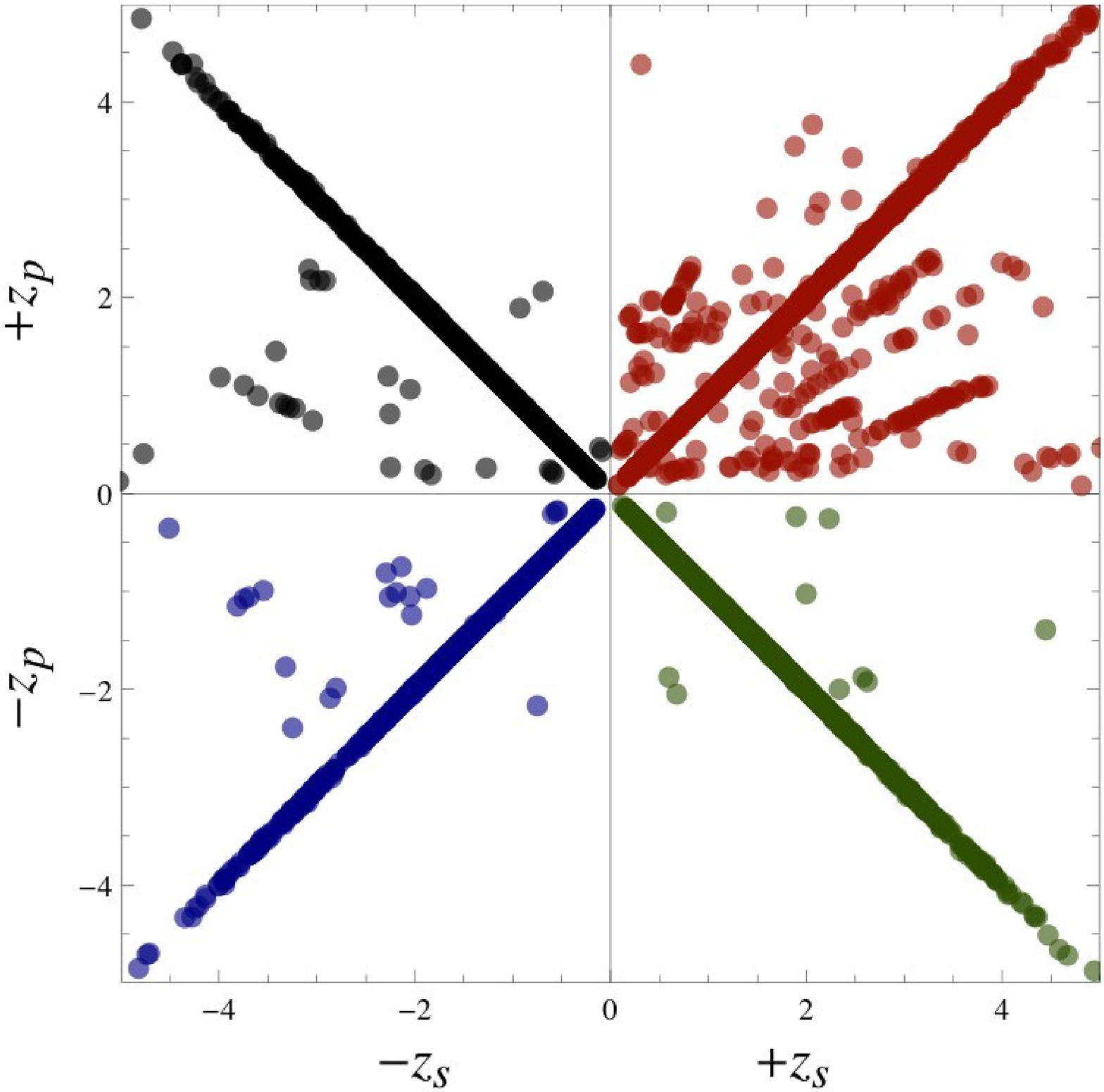}
\hspace{0.5cm}
\includegraphics[width=75mm]{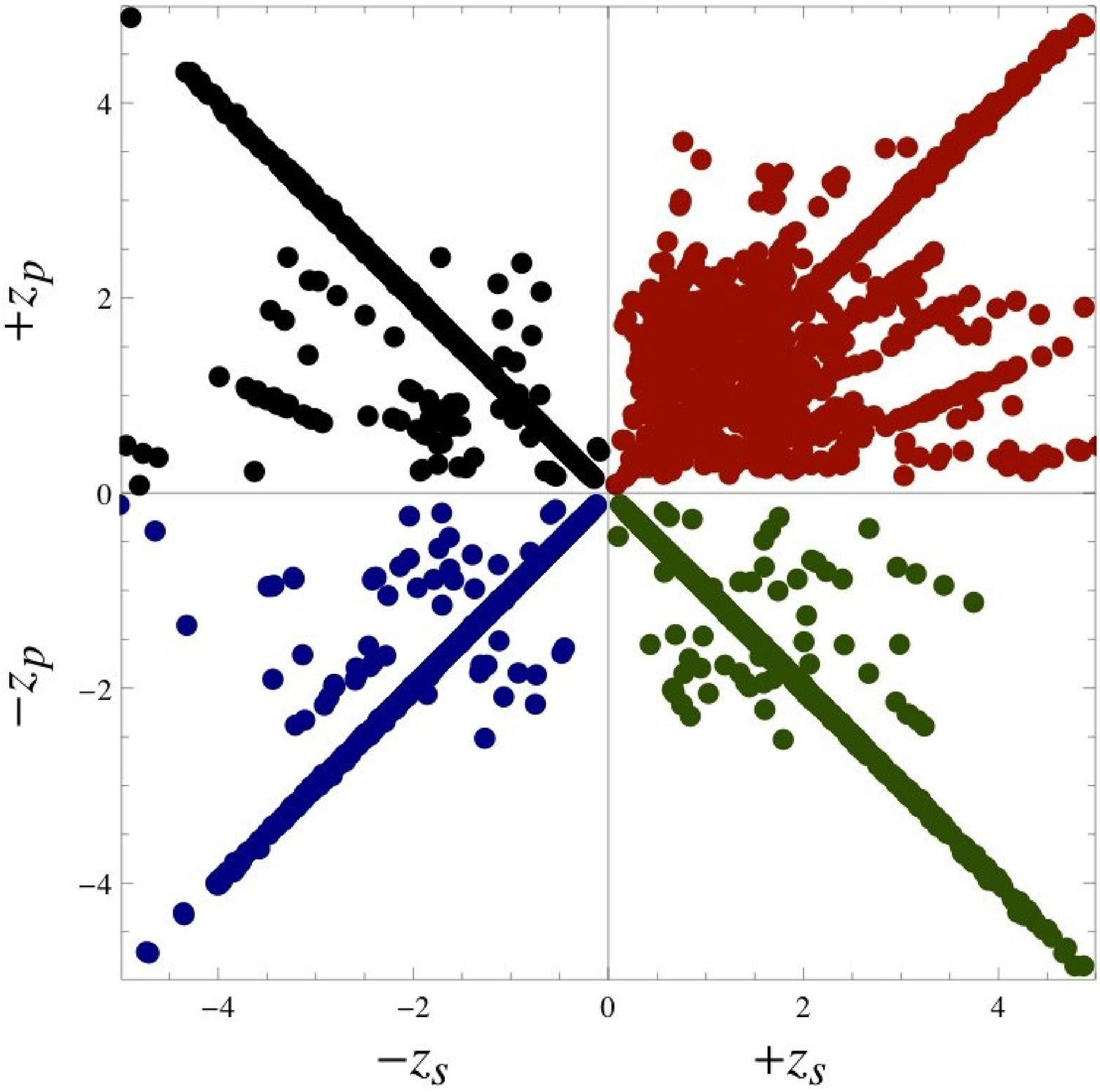}
\caption{
Scatter-plots of spectroscopic redshifts (horizontal axis)
{\it versus} photometric redshifts (vertical axis) obtained with the training set method,
for the four groups of decreasing sparseness (1, 2, 3 and 4, in decreasing sparseness). 
Left panel: original sample; 
right panel: simulated sample. As before, there are 2,500 objects in the first group 
(first quadrant in the upper right corner, red dots in color version); 
2,500 objects in the second group (second quadrant and green dots); 
2,500 objects in the third group (third quadrant and blue dots); and 2,500 objects in the
fourth group (fourth quadrant and black dots).
}
\label{Fig:scatterplot2}
\end{figure*}

The results
for the median and median deviation of $|z_s-z_p|/(1+z_s)$ are shown in 
Fig. \ref{Fig:medians_ts}. Although the fraction of outliers for groups 2-4 is 
roughly the same (at the level of 2-3\%), the median and the median deviation 
of the photo-z errors are clearly correlated with the density of the training set.
Comparing with Fig. \ref{Fig:medians} we see that the training set has a lower 
accuracy than the template fitting method -- both the median and the
median deviation of the training set groups are about twice as large as those 
of the template fitting groups.

The rms error after removing catastrophic objects with $\delta_z>0.02(1+z)$ 
is, for the original sample, 
$\sigma_z^{nc}/(1+z)=0.035$, 0.001, 0.0016 and 0.0037 for the sparseness
bins 1-4. For the simulated sample the rms errors after eliminating the outliers are
$\sigma_z^{nc}/(1+z)=0.082$, 0.0045, 0.0045 and 0.007 for the sparseness
bins 1-4.
For the photo-z groups 2, 3 and 4, the errors as measured by this criterium are 
about 2-3 times as large as the ones obtained with the template fitting 
method (see Fig. \ref{Fig:sigmas}).

\begin{figure*}
%\vspace{0.5cm}
\includegraphics[width=75mm]{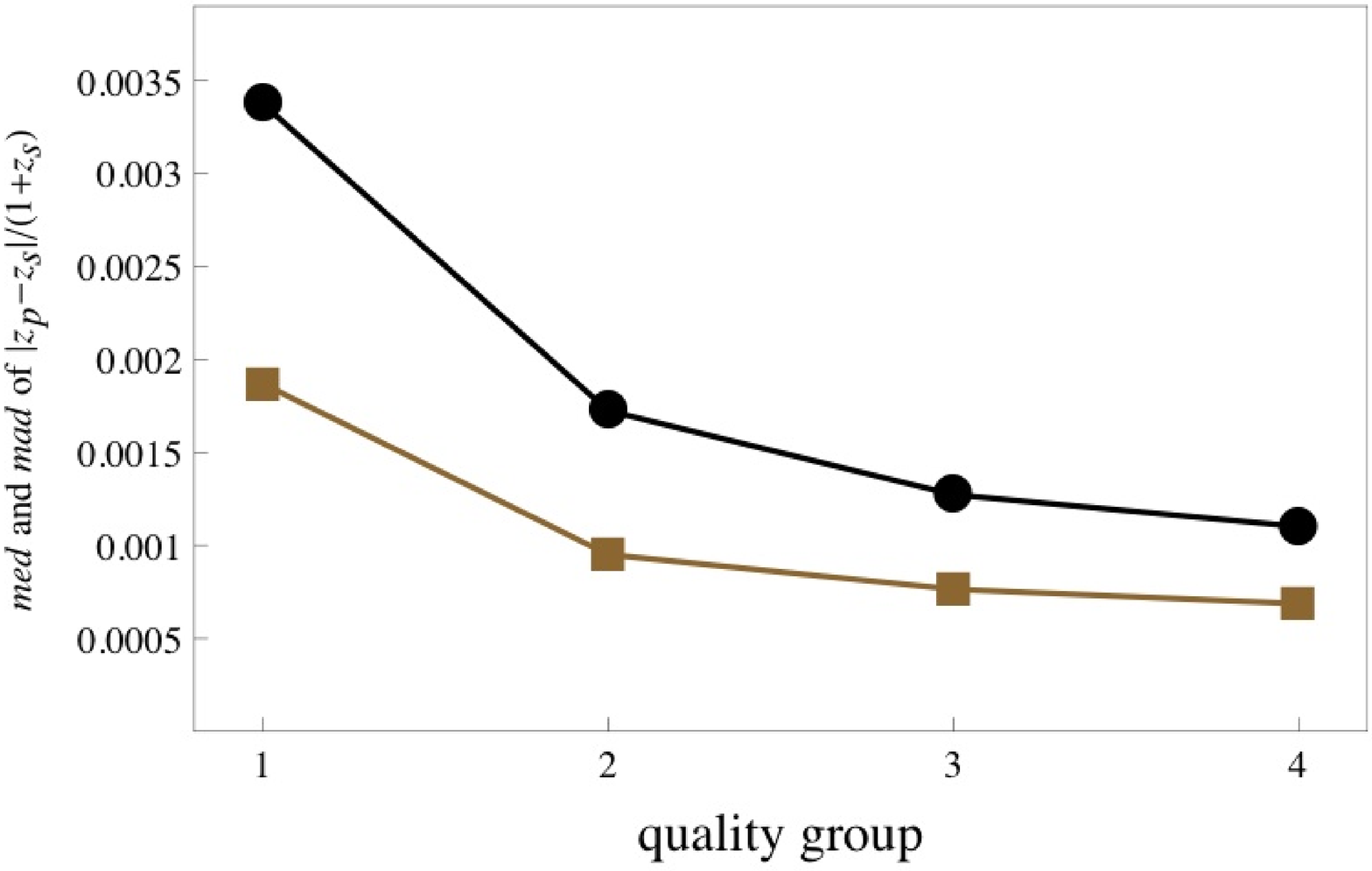}
\hspace{0.5cm}
\includegraphics[width=75mm]{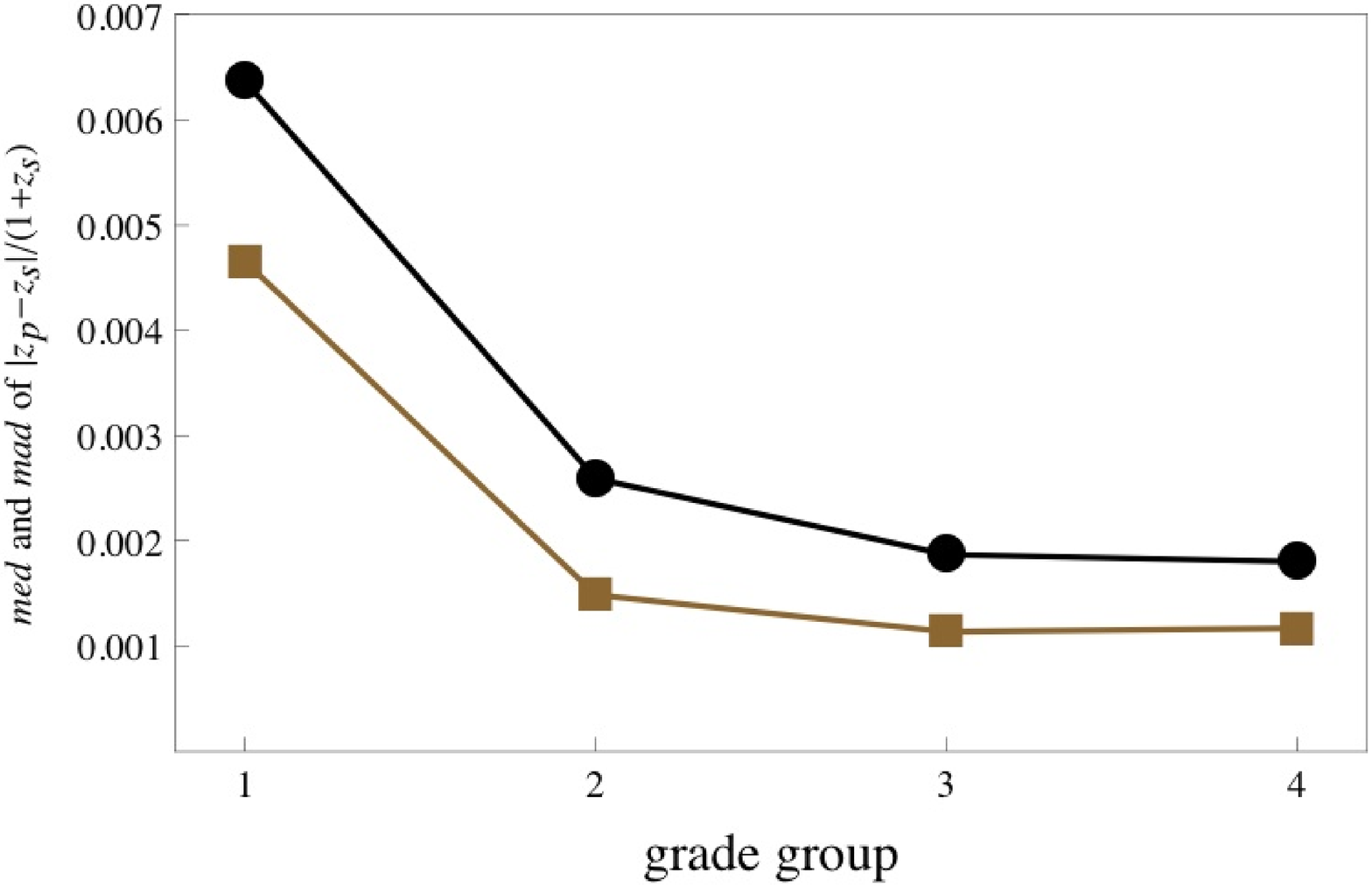}
\caption{Median ({\it med}) and median absolute deviation ({\it mad}) of the
errors in the photometric redshifts for the training set method. 
Left panel: original sample of SDSS quasars; 
right panel: simulated sample. The circles (black in color version) denote
the medians for each grade group; squares (brown in color version) denote
the {\it mad}.
}
\label{Fig:medians_ts}
\end{figure*}

We expect these results to improve significantly if we employ a denser training set. 
With the relatively sparse training set used here, we do not expect 
complex empirical methods to improve the photo-z accuracy. 
For instance, we have tried to use the set of the few nearest neighbors 
of a given object to fit a polynomial relation between 
magnitudes and redshifts, which we then applied to 
estimate the redshift of the photometric quasar. The results of such 
procedure were similar but slightly worse than simply taking the redshift
of the nearest neighbor. 
That happens because our quasar sample 
is not dense enough to allow for stable global -- and even local -- fits.
%Therefore, even though our results reflect the use of a very simple method, 
%we expect  to improve significantly with a larger and denser training set. 

With a sufficiently large training set, it has been shown that global neural 
network fits produce photo-z's of similar accuracy to those obtained 
by local polynomial fits [\citet{Oyaizu_2008}]. However these used a few hundred 
thousand training set galaxies spanning a redshift range of [0,0.3] whereas 
here we have $10^4$ quasars spanning the redshift range [0,5].

\subsection{Comparison of the template fitting and training set methods}

We have seen that the two methods for extracting the redshift of quasars, given a
low-resolution spectrum, yield errors of the same order of magnitude. 
Both the template fitting (TF) and the training set (TS) methods also yield empirical criteria for
selection of potential catastrophic redshift errors (the ``quality factor" of the
photo-z, in the case of the TF method, and the distance between 
nearest neighbors in the case of the TS method), which allows
one to improve purity at the price of reducing completeness.

%The accuracy of the template fitting method, after culling the catastrophic errors, is
%$\sigma^{nc}_z = 0.0022(1+z)$ and $\sigma^{nc}_z = 0.0025(1+z)$ for the original
%and simulated samples, respectively. The accuracy of the training set is similar:
%$\sigma^{nc}_z = 0.003(1+z)$ and $\sigma^{nc}_z = 0.004(1+z)$ for the original
%and simulated samples.

A larger sample of objects (the entire SDSS spectroscopic catalog
of quasars, for instance, has $\sim 10^5$ objects, instead of the $\sim 10^4$ that we 
used in this work) would improve the performance of the TS
method significantly, but may not necessarily make the performance of the TF 
method much better. A larger sample means a denser training set, which will certainly
lead to better matches between nearby objects, as well as a better overall accuracy. 
From the perspective of the TF method, a larger sample only means
a larger calibration set, and with our sample the performance of the method
is already being driven not by the calibration, but by intrinsic spectral variations in 
quasars -- something that the TS method is perhaps better suited to detect.

We have also applied a hybrid method to improve the quality of 
the photo-z's even further, by combining the power of the TF and TS methods
in such a way that one serves to calibrate the other.
The method was implemented for the simulated sample of quasars in the following manner.
First, we eliminate the 10\% worst photo-z's from the samples of quasars, either by
using the quality factor, in the case of the TF method, or by using the 
distance between nearest neighbors, in the case of the TS method.
This procedure alone reduces the median of the errors, $\Delta z/(1+z)$,
%to 0.001 (tf) and 0.0015 (ts) for the original sample, and 
to 0.0014 (TF) and 0.0024 (TS), and reduces
the fraction of outliers  
%to 1\% (tf) and 0.7\% (ts) for the original sample, and 
to 5\% (TF) and 4\% (TS).

The next step is to flag as potential outliers all objects which have been rejected
by either one of the 10\% cuts, and to eliminate them from both samples -- i.e., objects rejected
by one method are also culled from the sample that survives the cut from the other
method. The result is a culled sample containing about 83.6\% of the initial $10^4$ objects.
In that sample, the fraction of outliers is further reduced to 3.5\% (TF) and 2.6\% (TS).
%for both the original sample (8119 objects) and the simulated (8091) sample of quasars.

The final step is to compare the two photo-z's in the culled sets and
flag those that differ by more than a certain threshold,
namely $|z_{TF}-z_{TS}|/[1+0.5 (z_{TF}+z_{TS})] = 0.02$. 
After removing the flagged objects we still retain about 80\% of the original sample
(8001 quasars), but the fraction of outliers falls dramatically, to 0.6\% (47 objects out of 8001).
The median error for this final sample is 0.0013 (TF) and 0.0023 (TS),
and the median deviation is 0.00084 (TF) and 0.0014 (TS).

Hence, the combination of the TF and TS methods can yield 80\%
completeness with 99.4\% purity, and quasar photo-z errors which are as good as
the spectroscopic ones.
The histogram in Fig. \ref{Fig:compare} illustrates how this hybrid method
is able to identify the outliers, and Table \ref{Table1} shows how the performance
of the photo-z estimation is enhanced by the successive cuts.
Although the TS method is slightly better than the TF method at identifying the outliers,
it is significantly worse in terms of the accuracy of the photo-z's. 
However, the performance of the TS method should improve with a larger 
(and therefore denser) training set.

\begin{figure*}
\vspace{0.5cm}
\includegraphics[width=75mm]{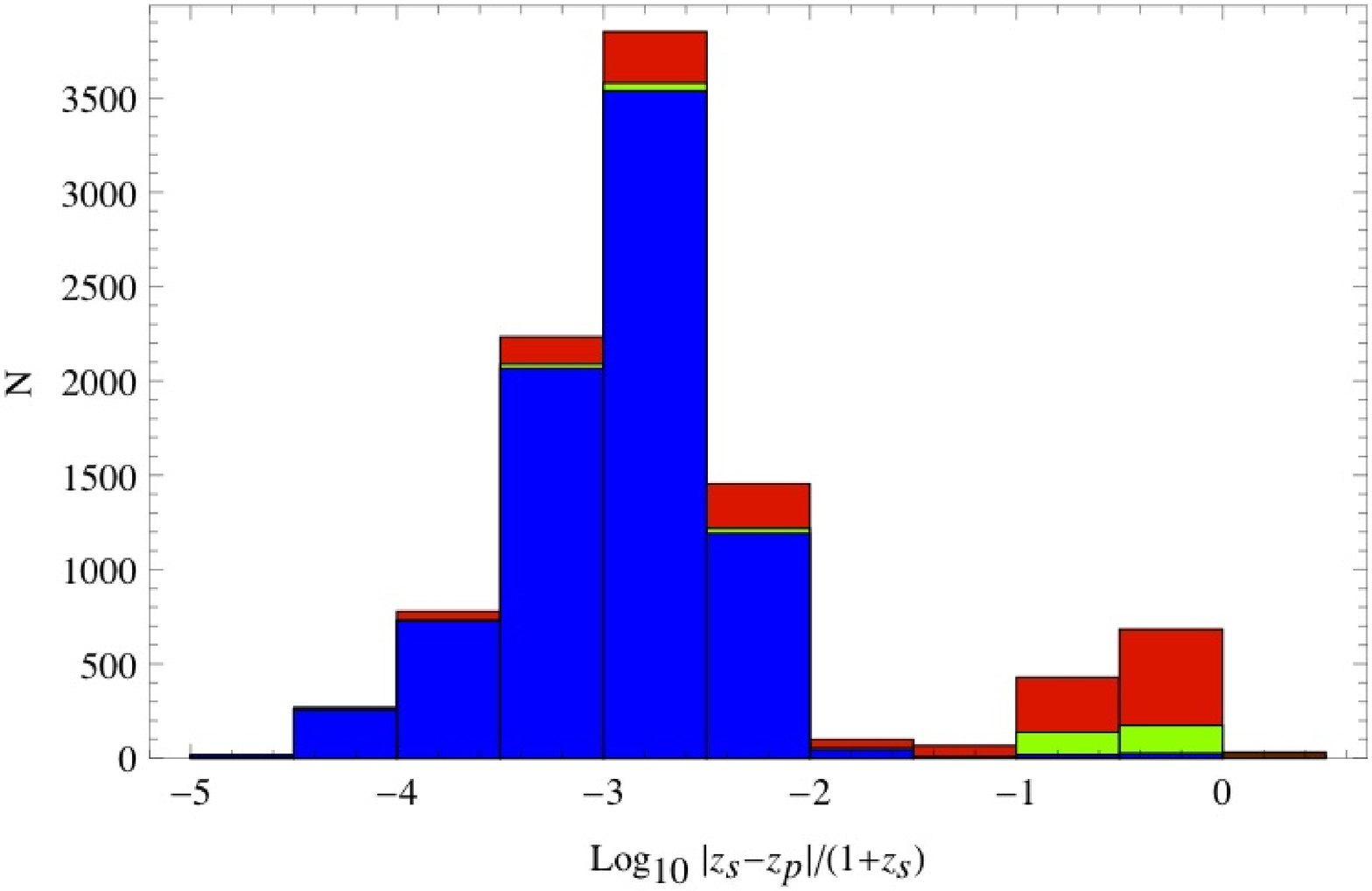}
\hspace{0.5cm}
\includegraphics[width=75mm]{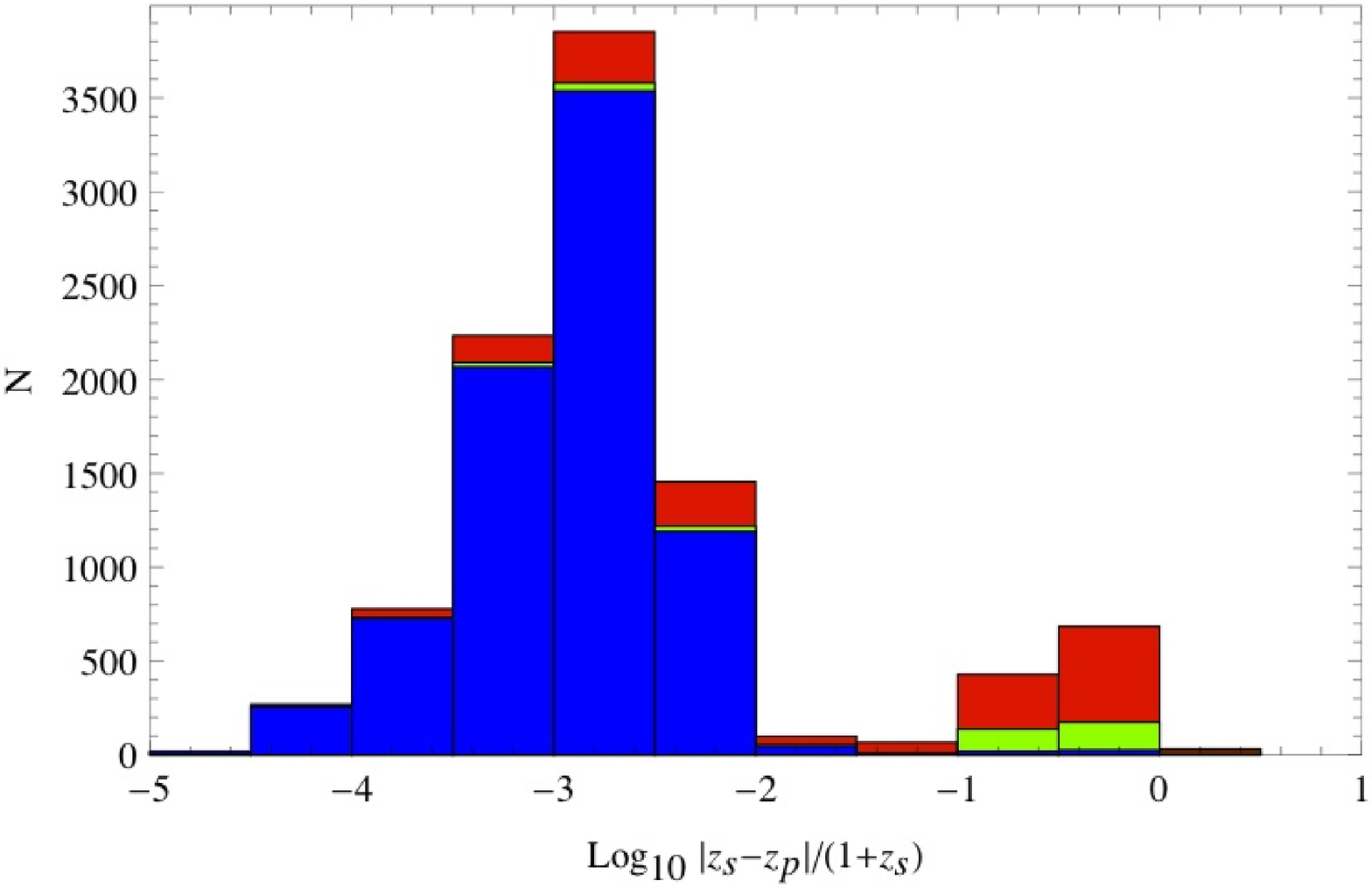}
\caption{Histograms of the photo-z errors for the simulated sample of quasars.
The left and right panels correspond to the template fitting (TF) and training set (TS) 
methods, respectively. The first quality cut (i.e., the quality factor in the case of
the TF method, and the distance between nearest neighbors 
in the case of the TS method) reduces the full sample of $10^4$ quasars 
by 15\% (upper bars, red in color version). 
The second cut, obtained by comparing the photo-z's from each method,
further reduces the number of quasars to 80\% of the full sample (i.e., 8001 objects).
The rate of outliers in this final sample is approximately 0.6\% -- see Table \ref{Table1}.
}
\label{Fig:compare}
\end{figure*}

%\begin{table}[htdp]
\begin{table*}
\caption{Completeness (fraction of objects that remain after applying the cuts),
purity (fraction of objects after culling the outliers) and accuracy
of the photo-z's for the simulated sample of quasars.
The first step eliminates the 10\% worst-quality photo-z's in both techniques, producing the
samples TF$_{90}$ and TS$_{90}$. The second step keeps only those objects
which are present both in TF$_{90}$ and in TS$_{90}$, producing the samples
TF$_{c}$ and TS$_{c}$. The last step is to compare the photo-z's that
were obtained with the different techniques, and to flag those that differ by more
than the threshold $\Delta z/(1+z) \geq 0.02$ as potential outliers.}
\label{Table1}
\begin{center}
\begin{tabular}{|c|c|c|c|}
\hline
Method & Completeness (\%) & Purity (\%) & $median \, [ \, \Delta z/(1+z) \, ]$ \\
\hline
TF$_{90}$ = TF - TF$_{10}$ & 90 & 95 &  0.0014 \\
TS$_{90}$ = TS - TS$_{10}$ & 90 & 96 &  0.0024 \\
\hline
TF$_{c}$ = TF$_{90}$ - TS$_{10}$ & 85 & 96.5 &  0.0014 \\
TS$_c$ = TS$_{90}$ - TF$_{10}$ & 86 & 97.4 &  0.0023 \\
\hline
TF$_c$ v. TS$_c$ & 80 & 99.4 &  0.0013 \\
TS$_c$ v. TF$_c$ & 80 & 99.4 &  0.0023 \\
\hline
\end{tabular}
\end{center}
\label{default}
\end{table*}%

As a final note, there are a few important factors that we have not considered,
which may affect the performance of the quasar photo-z's. One of them
is the calibration of the filters, which, if poorly determined, could introduce
fluctuations of (typically) a few percent in the fluxes. Since J-PAS uses a secondary, 0.8 m
aperture telescope dedicated to the calibration of the filter system, the stated 
goal of reaching 3\% global homogeneous calibration seems feasible -- and, 
in fact, we employed that lower limit for the noise level of our simulated quasar sample.
An even more important factor is the time variability of the intrinsic SEDs of quasars, 
which can be a much larger effect than the fluctuations induced by calibration errors.
Since a final decision concerning the strategy of the J-PAS survey has not yet been 
reached at the time this paper was finished, we decided not to pursue a simulation 
that took variability into account.
However, it seems likely that each quasar that is observed by J-PAS will have
several (7 or more) adjacent filters measured during an interval of a few (4-10) days, 
at most, and the full SED will be represented by a few (4-8) of these snapshots.
In that sense, the information in the time domain contained by these snapshots would 
not be simply a nuisance, but may be used to aid in the identification of the 
quasars.

%% ADDED SUBSECTION
\subsection{Completeness and contamination}

%\begin{figure*}
%\vspace{0.5cm}
%\includegraphics[width=80mm]{Quasars_Stars_chi2_ratio}
%\caption{Segregation of stars (green squares), quasars with catastrophic photo-z's (red diamonds)
%and quasars with good photo-z's (blue circles) in terms of the value of $\chi^2$ (vertical
%axis) and $r$ (see Section 2.3). For this plot we randomly selected 2000 stars, 1600 quasars with good %photo-z's, and 400 quasars with poor photo-z's.
%}
%\label{Fig:contamination}
%\end{figure*}

In order to understand how a quasar sample produced from an optical narrow-band
survey could be contaminated by other types of objects (stars, mostly), 
we have used data from SDSS spectroscopic plates in which a random subsample 
of all point sources with $i < 19$ had their spectra taken
[\citet{AdelmanMcCarthy:2005se}]. We randomly extracted $10^4$ 
stars from this catalog, and processed their spectra using the template fitting method
that was outlined in the previous subsections for the SDSS simulated quasars.
However, we do not include any star templates in our fitting procedure, so the
only questions we are asking are: ({\it i}) what is the redshift which best fits the quasar template 
spectrum to the spectra of each individual star, and ({\it ii}) what are the qualities of those 
fits (their reduced $\chi^2$)?

\begin{figure*}
\vspace{0.5cm}
\includegraphics[width=76mm]{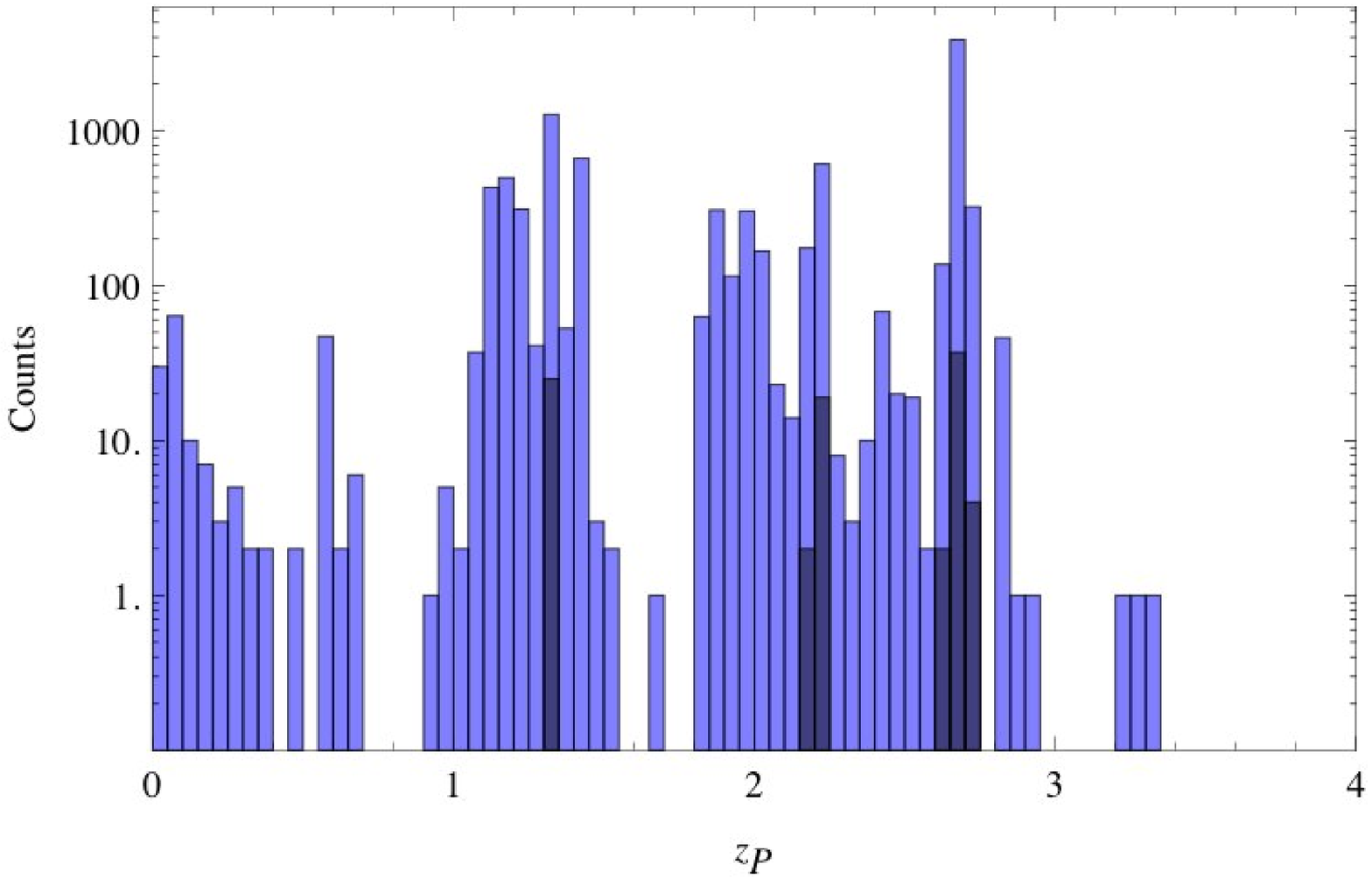}
\hspace{0.5cm}
\includegraphics[width=75mm]{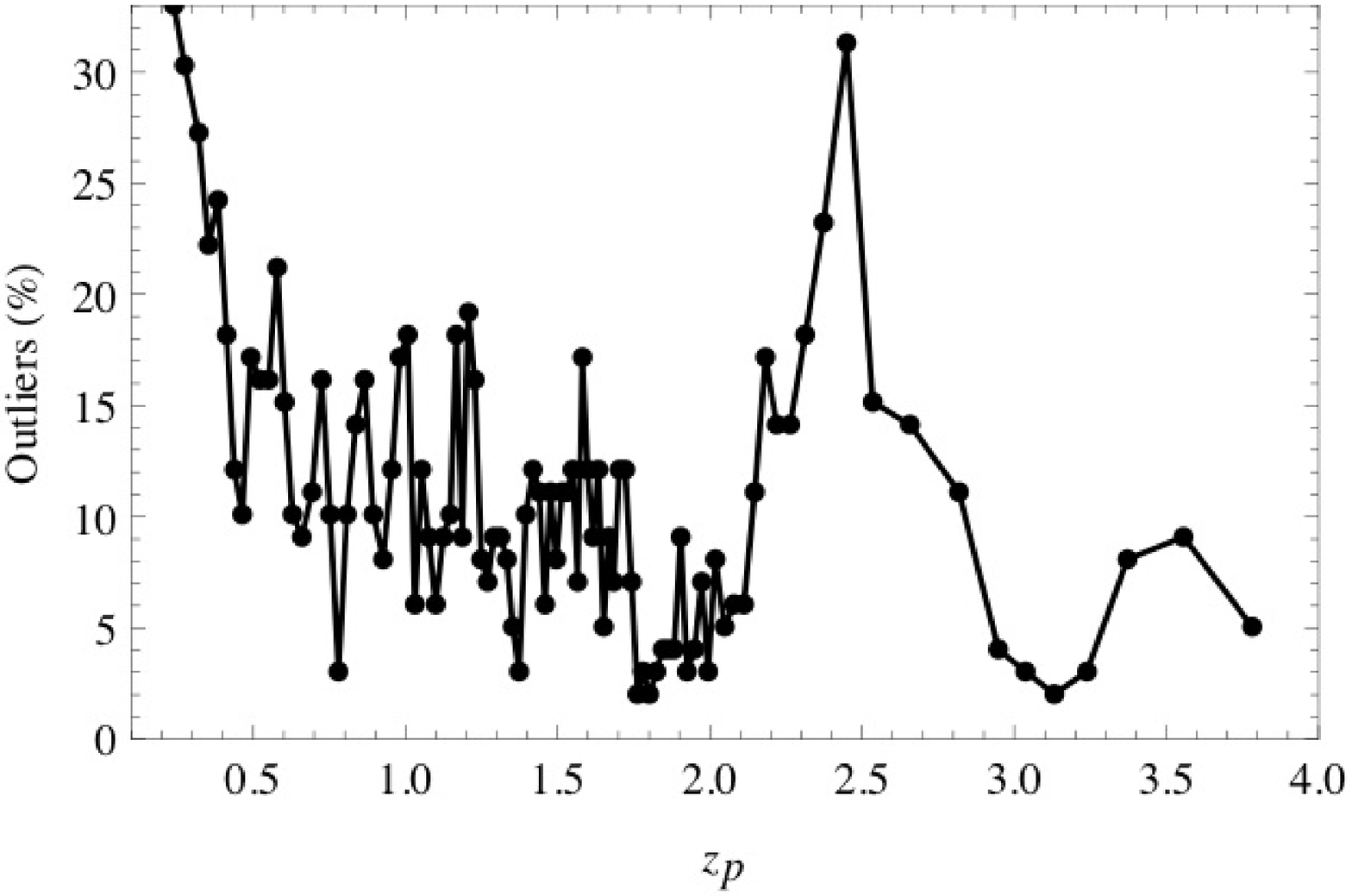}
\caption{Left panel: photo-z's assigned to stars by the quasar template fitting
code, in 100 bins of redshift. The 1\% stars with the highest potential to be 
confused with quasars (i.e., those whose spectra satisfy both $\chi^2 < 3.0$ and $r<0.75$)
are shown as dark bars.
Right panel: fraction of photo-z outliers [i.e., quasars whose
photo-z's differ from the correct redshifts by more than 0.02(1+z)] for the simulated quasar catalog, 
in 100 bins of redshift.
}
\label{Fig:zstars}
\end{figure*}

Using the tools which were introduced in Section 2.3 we were able to 
reject the overwhelming majority of stars, just on the basis of their poor 
reduced $\chi^2$ fits to the quasar template, and the degeneracy in their photo-z's 
as measured by the parameter $r$ of Section 2.3
(stars lack the quasar's emission-line features, which are the
key determinants of the photo-z's, and this translates into high values of $r$). 
Hence, it is clear that, in this sense, 
stars are quite segregated from quasars -- and the introduction of stellar templates 
would further improve this separation. As a comparison,
the COMBO-17 quasar catalog [\citet{Wolf_COMBO17_q}] doesn't suffer from 
significant contamination from stars, even though it has a lower spectral resolution
than J-PAS, and similar depths. Hence, we conclude that the prospects of 
J-PAS achieving high levels of purity and completeness are quite good -- however, 
we cannot definitively answer this question here, and leave this critical issue to 
future work.

Nevertheless, we can determine which redshift ranges are most likely to affect the 
completeness and purity of the quasar sample due to contamination from stars. 
Fig. \ref{Fig:zstars} shows that the photo-z's falsely assigned 
to stars are concentrated in a few intervals, corresponding to redshifts where
the visible region of the quasar spectra present few distinguishing features. 
The concentration of false photo-z's in narrow intervals is starker for those stars 
whose spectra can most easily be confused with those of quasars -- which, for the
purposes of this exercise, are stars whose fits to the quasar template 
satisfy both $\chi^2<3$ and $r<0.75$ (approximately 1\% of the total). 
Some of these problematic redshift intervals also contain a large 
proportion of the catastrophic photo-z's for the true quasars (see the right panel of
Fig. \ref{Fig:zstars}). These plots indicate that contamination should be a greater 
concern for the redshift intervals $1.30 \lesssim z_p \lesssim 1.31$, 
$2.2 \lesssim z_p \lesssim 2.22$ and $2.65 \lesssim z_p \lesssim 2.7$.

%%%%%%%%%%%%%%%%%%%%%%%%%%%%%%%%%%%%

\section{Quasars as cosmological probes}

The SDSS sample of quasars [\citet{richards_photometric_2001,vanden_berk_composite_2001,schneider_sloan_2003,yip_spectral_2004,schneider_sloan_2007,shen_clustering_2007,ross_clustering_2009-1,schneider_sloan_2010,shen_catalog_2010}]
has enabled a reliable measurement of the quasar luminosity function 
[\citet{richards_2df-sdss_2005,richards_sloan_2006,2007ApJ...654..731H,croom_2slaq,croom_2df-sdss_2009}], which, in terms of the g-band absolute magnitude is given by the fit
[\citet{croom_2df-sdss_2009}]:
\be
\label{Eq:LumFun}
\phi(M_G,z) = \frac{\phi_0}{10^{0.4 \, (1+ \alpha) \, [M_G-M_G^*(z)]} + 
10^{0.4 \, (1+ \beta) \,  [M_G-M_G^*(z)]}} \; ,
\ee
where $\phi_0 = 1.57 \times 10^{-6}$ Mpc$^{-3}$,
$\alpha = -3.33$, $\beta = -1.41$ and the break magnitude 
expressed in terms of $M_G$ is given by:
\be
\label{MBreak}
M_G^*(z) = -22.2 - 2.5 \, (1.44 \, z - 0.32 \, z^2) \; .
\ee 
Notice that the quasar luminosity function and the break magnitude were 
obtained with a sample of quasars only up to $z\sim 2.5$, and it is far from
clear that these fits can be extrapolated to higher redshifts and lower 
luminosities [\citet{croom_2df-sdss_2009}].

To obtain the number density of quasars as a function of some
limiting (absolute) magnitude $M_G^0$, the luminosity function 
above must be integrated up to that magnitude.
In Fig. \ref{Fig:LumFun} we plot the quasar volumetric density both
in terms of the limiting apparent magnitude in the $g$ band for flux-limited
surveys, $n(<g_{lim})=\int_{-\infty}^{g_{\lim}} dg \ \phi(g)$ 
(solid lines, $g_{\lim} =$ 24, 23, 21 and 19, 
from top to bottom), and also in terms of the absolute 
magnitudes $n(<M_{G , \lim})=\int_{-\infty}^{M_{G , \lim}} dM_G \ \phi(M_G)$
(dashed lines, $M_{G , \lim} = $ -20, -22, -24 and -26 from
top to bottom.) 
Since contamination from the host galaxy may hinder our ability to identify
low-luminosity quasars through color selection (this can be especially problematic
at low redshifts), we chose to apply a cut in absolute magnitude in the luminosity
function, in addition to the apparent magnitude cut. 

As a concrete example, we will discuss a flux-limited survey up to an apparent
magnitude of $g<23$, and include only those objects which are more luminous 
than $M_G<-22$, since quasars fainter than this often have their light dominated
by the host galaxy.
The resulting comoving number density is shown
as the dashed line and hashed region in Fig. \ref{Fig:LumFun}, which peaks
at $z\sim 1.6$ with $n_{\rm max} \sim 10^{-5}$ Mpc$^{-3}$
(or $\sim 3. 10^{-4} \, h^3$ Mpc$^{-3}$.) If the limiting 
apparent magnitude is $g<24$, the number density can be as large as
$10^{-4} \, h^3$ Mpc$^{-3}$ at $z \sim 2$.
As we will see below, the relatively small density of quasars when compared 
to galaxies (which can easily reach $n \gtrsim 10^{-3}$ Mpc$^{-3}$) is 
compensated by the facts that quasars are highly biased 
tracers of large-scale structure, and that the volume that they span is
larger than that which can be achieved with galaxies -- for a similar analysis, 
see also \citet{Wang_LAMOST,2011arXiv1108.1198S}.

\begin{figure}
\vspace{0.5cm}
\includegraphics[width=80mm]{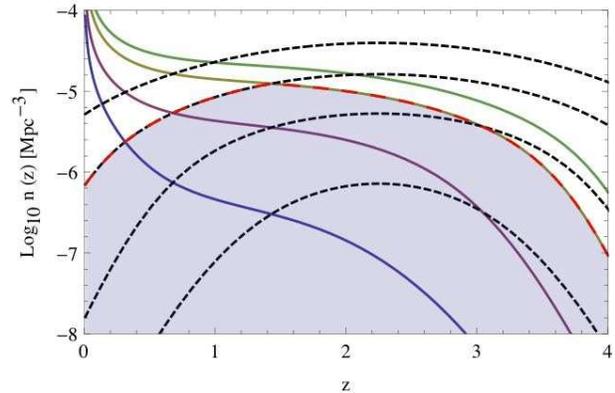}
\caption{
The volumetric density of quasars for different
limiting $g$-band apparent magnitudes (solid lines) 
and different absolute magnitudes (dashed lines), as a function of
redshift, computed according to the luminosity function of Croom {\it et al.} 2009.
The solid lines, from top to bottom, 
correspond to limiting magnitudes of
$g \leq$ 24 (green in color version), 23 (yellow in color version), 
21 (red in color version) and 19 (blue in color version); 
the short-dashed lines, from top to bottom, correspond to 
absolute luminosity cut-offs of
$M_G \leq$ -20, -22, -24 and -26 respectively.
}
\label{Fig:LumFun}
\end{figure}

It is also useful to compute the total number of quasars
that a large-area (1/5 of the sky), flux-limited survey could 
produce -- assuming the quasar selection is perfect. 
In Fig. \ref{Fig:QSONumbers} we show that an $8.4 \times 10^3$ deg$^2$
survey up to $g<23$ ($g <24$) could yield $2.0 \times 10^6$ 
($3.0 \times 10^6$) objects, up to $z = 5$.

\begin{figure}
\vspace{0.5cm}
\includegraphics[width=80mm]{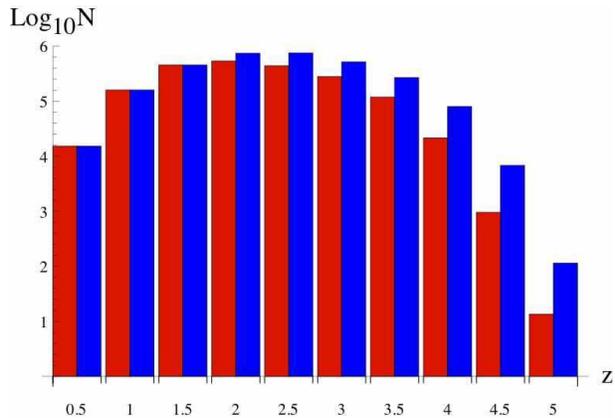}
\caption{
Total numbers of quasars in $\Delta z=0.5$ bins for an
$8.4 \times 10^3$ deg$^2$ survey, assuming a $5\sigma$ 
point-source magnitude limit of $g=$ 23 (left bars, red in
color version) and 24 (right bars, blue in color version.) The
numbers are identical for $z \leq 1.5$ because our selection criteria
culls the quasars fainter than $M_G = -22$, which means that
for $z<1.5$ the catalog is equivalent to a volume-limited 
and absolute magnitude-limited survey.
}
\label{Fig:QSONumbers}
\end{figure}

\subsection{Large-scale structure with quasars}

Quasars, like any other type of extragalactic sources, are
biased tracers of the underlying mass distribution:
$P_{q} (k,z) = b_{q}^2(z) P(k,z)$, where $P(k,z)$ is the
matter power spectrum, $P_{q} (k,z)$ is quasar power spectrum
(the Fourier transform of the quasar two-point correlation function), 
and $b_{q}$ is the quasar bias. The quasar bias is a steep function of redshift 
[\citet{shen_clustering_2007,ross_clustering_2009-1}], 
and it may depend weakly on the intrinsic (absolute)
luminosities of the quasars [\citet{lidz_luminosity_2006}], but it is thought
to be independent of scale ($k$) -- at least on large scales.

The connection between theory and observations is further
complicated by the fact that both the observed two-point 
correlation function and the power spectrum inherit an 
anisotropic component due to redshift-space distortions
[\citet{hamilton_linear_1997}].
In this work we will only consider the {\it monopole}  of 
the power spectrum, $P(k) = \int_{-1}^1 d\mu P(k,\mu)$, where 
$\mu$ is the cosine of the angle between the tangential and the radial 
modes. We will address the full redshift-space
dataset from our putative quasar survey, as well as the 
resulting constraints thereof, in future work. 
%Since inclusion of 
%the directional information can only add to the information in the monopole, 
%the present work can be regarded a conservative 
%lower bound on the power of quasars to constrain large-scale structure.

To first approximation the statistical uncertainty in the
power spectrum can be estimated using the formula 
derived in \citet{feldman_power-spectrum_1994} for three-dimensional surveys:
\be
\label{Eq:StatErrPk}
\frac{\Delta P(k,z)}{P(k,z)} \simeq \sqrt{\frac{2}{N_{m}(k,z)}}
\left[ 1 + \frac{1}{n(z) b^2 (z) P(k,z)} \right] \; ,
\ee
where $n$ is the average number density of the objects used to trace 
large-scale structure, and $b$ is the bias of that tracer.
The number of modes (the statistically independent 
degrees of freedom) in a given bin in $k$-space is given by
$N_{m} = 4 \pi V(z,z+\Delta z) k^2 \Delta k /(2\pi)^3$,
where $\Delta z$ and $\Delta k$ denote the thickness of the 
redshift bins and of the wavenumber bins, respectively. 
The first term inside the brackets in Eq. \ref{Eq:StatErrPk} corresponds 
to sample variance, and the second corresponds to shot noise (assuming the
variance of the shot noise term is that of a Poisson distribution of the counts.)
Since the power spectrum peaks at $P \lesssim 10^{4.5} \, h^{-3}$ Mpc$^3$, 
a quasar survey with $n \lesssim 10^{-5} \, h^3$ Mpc$^{-3}$ would be 
almost always limited by shot noise.

For the purposes of this exercise we have used 28 bins in Fourier space,
equally spaced in $\log(k)$, and spanning the interval
between $0.007$ $h$ Mpc$^{-1} < k < 1.4$ $h$ Mpc$^{-1}$.
Our reference matter power spectrum $P_0(k,z)$ is  
a modified BBKS spectrum [\citet{bardeen_statistics_1986}]
[see also \citet{Peacock} or \citet{amendola_dark_2010}].
The transfer function of the BBKS fit does not contain the
BAO modulations, so we have modeled those features in
the spectrum by means of the fit [\citet{Seo:2007ns}; see also \citet{Benitez:2008fs}]:
\be
\label{Eq:PkBAO}
P(k,z) = P_{0}(k,z) 
%\\ \nonumber
%& \times &
\left[1 + k A \sin (k r_{BAO}) e^{ - k^2 R^2} \right]
\; ,
\ee
where $r_{BAO} = 146.8$ Mpc = $105.7 \, h^{-1}$ Mpc  is the
length scale of the BAOs that can be inferred from WMAP 
[\citet{Hinshaw:2008kr}], $A=0.017 \, r_{BAO}$ is the amplitude
of the acoustic oscillations, and $R = 10 \, h^{-1}$ Mpc denotes the Silk 
damping scale.

Eq. (\ref{Eq:StatErrPk}) is an approximation which is
appropriate for spectroscopic redshift surveys, although this is not 
the type of survey that we are considering. Nevertheless, we have showed in
the previous Section that, with narrow-band filters, the error in the photo-z's 
of quasars can be lower than $\delta z \sim 0.002 \, (1+z)$,
which is excellent but not quite equivalent to a spectroscopic
redshift. Redshift errors smear structures on small scales along the
line-of-sight, and can be factored into the estimation of the
power spectrum through an empirical damping term [\citet{angulo_detectability_2007}]:
\be
\label{Eq:z_k}
\exp{ \left[ - k_\parallel^2 \frac{ \delta_z^2 c^2}{H^{2}(z)} \right] } \; ,
\ee
where $k_\parallel = k \mu$ denote the modes along the line-of-sight.
In our $\Lambda$CDM model, photometric redshift errors 
suppress modes which are smaller than about 
$k_{\parallel}^{-1} \, \sim \, \delta_z \times 10^3 h^{-1} $ Mpc at $z=2$ 
(or $k_{\parallel}^{-1} \, \sim \, \delta_z \times 5.10^2 h^{-1} $ Mpc at $z=4$). 
Hence, a quasar photo-z error of the order of $0.002 (1+z)$ only starts to affect
the power spectrum at scales $k_{\parallel} \gtrsim 0.2 h$ Mpc$^{-1}$ at z=2, and 
$k_{\parallel} \gtrsim 0.4 h$ Mpc$^{-1}$ at z=4. This is smaller 
than either the Silk damping scale or the scales at which nonlinear effects 
kick in (see the discussion below), so we expect that photo-z errors will be a 
subdominant nuisance in the estimation of the power spectrum and derived 
parameter constraints.

Another important point concerning Eq. (\ref{Eq:StatErrPk}) is that it
applies to the power spectrum as estimated by some biased tracer, but
it does not automatically include the uncertainty in the bias or the
selection function, or other systematic effects such as bias
stochasticity [\citet{dekel_stochastic_1998}]. 
Here we employ the fit found by [\citet{ross_clustering_2009-1}] 
for quasars with $z < 2.2$, which is given by \hbox{$b_q(z)= 0.53 + 0.29 \, (1+z)^2$.}
Although this bias has large uncertainties, especially at high redshifts,
we will implicitly assume that $b_{q} (z)$ is a linear, deterministic bias 
that has been fixed at each redshift
by this fit.

\begin{figure}
\vspace{0.5cm}
\includegraphics[width=80mm]{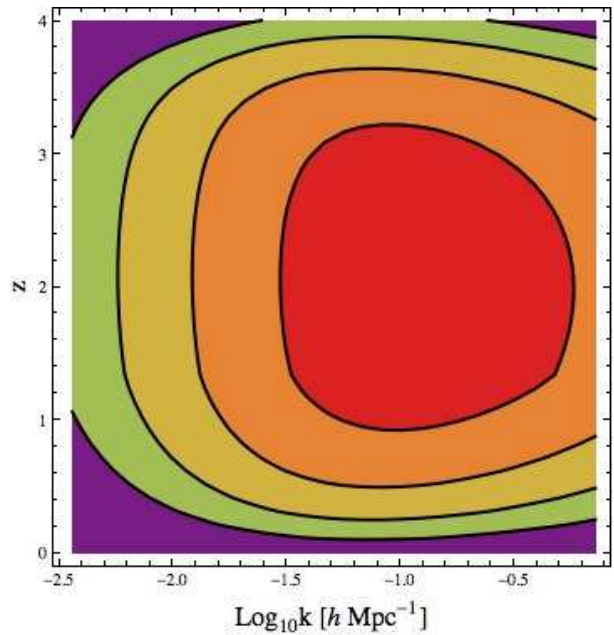}
\caption{The contours denote the 
statistical errors in the power spectrum, 
$\log_{10} \Delta P(k,z)/P(k,z)$, for an $8.4 \times 10^3$ deg$^2$ quasar 
survey, flux-limited down to $g < 23$, and limited to objects brighter than $M_G=-22$. 
From inside to outside, the contours correspond to $\Delta P/P = 10^{-1.5}$, 
$10^{-1}$, $10^{-0.5}$ and $10^{0}$. 
The uncertainties were computed using Eq. \ref{Eq:StatErrPk}.
For this plot we binned the redshift slices in intervals of 
$\Delta z =0.1$, and the wavenumbers were divided into
28 equally spaced bins in $\log(k)$, spanning the interval
between $k=0.007 \, h$ Mpc$^{-1}$ and $k=1.4 \, h$ Mpc$^{-1}$. 
Photo-z errors and uncertainties in
the bias of quasars are not included in our error budget.
}
\label{Fig:PkErrors}
\end{figure}

%We can now proceed to study the statistics of a survey of quasars such 
%as that which will be produced by J-PAS.
In Fig. \ref{Fig:PkErrors} we plot the contours corresponding
to equal uncertainties in the power spectrum as a function of
the scale [$\log_{10} k$ (h Mpc$^{-1}$), horizontal axis)] and redshift $z$ 
(vertical axis), according to Eqs. (\ref{Eq:LumFun})-(\ref{Eq:PkBAO}), 
and assuming that the J-PAS survey covers $8.4 \times 10^3$ deg$^2$ to a 
$5\sigma$ limiting magnitude of $g<23$. 
There are three main effects that determine the shape of the contours in Fig. 
\ref{Fig:PkErrors}: first,
at fixed $k$ and low redshifts, both the volume of the survey
as well as the number density of objects (which is determined by the absolute
luminosity cut) are small, while at high redshifts the number 
density falls rapidly due to the apparent magnitude cut.
Second, for a fixed $z$ the uncertainty as a function 
of $k$ decreases up to scales $k \sim 0.02 \, h$ Mpc$^{-1}$, 
where $P(k)$ peaks, and as it starts to fall, it increases the 
Poisson noise term in Eq. (\ref{Eq:StatErrPk}). Finally, the redshift 
evolution of the power spectrum [$P(k,z) \sim D^2(z)$, where $D(z)$ is the linear
growth function] also increases the shot noise at higher redshifts 
-- although this effect is partly mitigated by the redshift evolution
of the quasar bias.
Quasars achieve their best performance in estimating the power
spectrum at $z\sim$ 1 -- 3. This is because in that range the quasar bias 
increases faster than the number density falls as a function of redshift.

A closely related way of assessing the potential of a survey to
measure the power spectrum is through the so-called effective volume:
$$
V_{\rm eff} (k) = \int d^3 x \left[ \frac{ n \, b^2 \, P(k)}{1+n \, b^2 \, P(k)} \right]^2 \, ,
$$
where $x$ is comoving distance, and both the average number density $n$ and
the bias $b$ are presumably only functions of $x$ (or, equivalently, of redshift).
The effective volume is simply (twice) the Fisher matrix element that corresponds to
the optimal (bias-weighted) estimator of the power spectrum 
[\citet{feldman_power-spectrum_1994,1998ApJ...499..555T}]. In Fig. \ref{Fig:EVol}
we show the effective volume for our quasar survey (full line). For comparison,
we have also plotted the effective volume of a hypothetical quasar survey similar
to BOSS or BigBOSS, that would target $\sim 5. 10^5$ objects 
over the same area and with the same redshift
distribution as the J-PAS quasar survey (long-dashed line). Also plotted in
Fig. \ref{Fig:EVol} are the effective volumes of two surveys of LRGs assuming
the luminosity function of \citet{2007ApJ...654..858B}, 
either in the case of a shallow survey flux-limited to $g<21.5$ 
(``SDSS-like'', short-dashed line), or for a deep survey limited to $i<23$ 
(``J-PAS-like'', dashed line.)

\begin{figure}
\vspace{0.5cm}
\includegraphics[width=80mm]{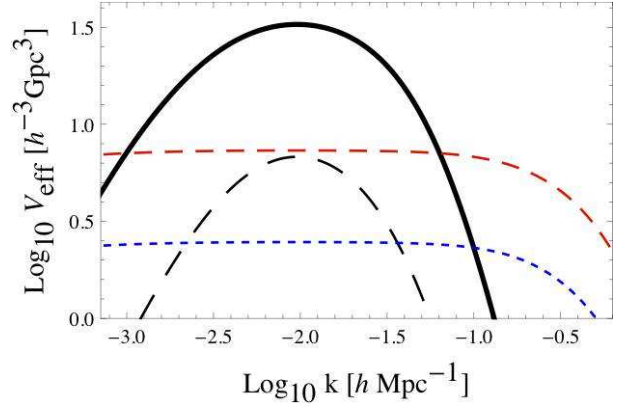}
\caption{
Effective volume of a flux-limited quasar catalog ($g<23$ and $z<4$) over 
$8.4 \times 10^3$ deg$^2$.
We also show the effective volume of a putative spectroscopic survey of quasars
with $4. 10^{5}$ objects, where we assumed the same area and redshift
distribution as was used for the J-PAS catalog (``BOSS-like", long-dashed line.)
For comparison, we also show two hypothetical catalogs of luminous 
red galaxies (LRGs) over the same area, one limited to $g<21.5$ (``SDSS-like'', 
short-dashed line, blue in color version) and the other limited 
to $i<23$ (long-dashed line, red in color version.) 
For the LRG estimates, we used the luminosity function of Brown {\it et al.} (2007)
and assumed a constant bias $b_{\rm LRG}=1.5$.
}
\label{Fig:EVol}
\end{figure}

%In Fig. \ref{Fig:3zs_errors} we plot the spectrum $P(k,z)$
%for $z=0.5$, 1.0, 1.5, 2.0 and 2.5, using redshift bins of 
%$\Delta z =0.1$ -- which means that there are four
%additional statistically independent sets of datapoints 
%that measure the power spectrum between each of the plotted
%curves. The uncertainties from Eq. \ref{Eq:StatErrPk} are
%denoted by the error bars for each bin in Fourier space. 

In Fig. \ref{Fig:3zs_baos_errors} we plot the power
spectrum divided by the BBKS power spectrum 
$P_0(k)$, in order to highlight the BAO features.
The error bars, from leftmost to rightmost (black in color version to orange in color version), 
corresponds to measurements of the power spectrum in redshit bins of $\Delta z=0.5$
centered in $z = 0.5$, $z = 1.0$, $z = 1.5$, $z = 2.0$, $z = 2.5$ and $z = 3.0$, respectively.
The power spectrum at low redshifts is 
poorly constrained, but this improves at high redshifts ($z \sim 1-3$).

%\begin{figure}
%\vspace{0.5cm}
%\includegraphics[width=80mm]{Pk_5zs_errors}
%\caption{
%Linear mass power spectrum with sample and Poisson noise for
%$z=0.5$ (top, black curve and error bars), $z=1.0$ (red), 
%$z=1.5$ (green), $z=2.0$ (blue), $z=2.5$ (purple) and
%$z=3.0$ (bottom, orange). For this figure we employed 
%redshift bins of $\Delta z = 0.1$ -- i.e., there are four additional
%statistically independent sets of points between each one of these
%curves.
%}
%\label{Fig:3zs_errors}
%\end{figure}

\begin{figure}
\vspace{0.5cm}
\includegraphics[width=80mm]{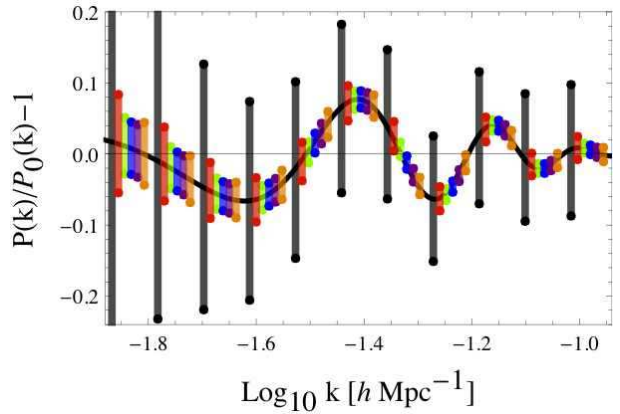}
\caption{
Baryon acoustic oscillations in position space. The oscillations are highlighted by 
dividing the full spectrum by a reference BBKS spectrum $P_{0}(k)$ 
without the baryon acoustic features.
From left to right, the error bars correspond to the uncertainties at 
$z=0.5$ (black curve and grey error bars), $z=1.0$ (red), $z=1.5$ (green), 
$z=2.0$ (blue), $z=2.5$ (purple), and
$z=3.0$ (orange). In this plot we employed redshift bins of $\Delta z= 0.5$.
The errors of the $z=0.5$ bin are much larger than those of other bins because:
i) the volume of the $z=0.5$ bin is much smaller than that of other bins, which
makes cosmic variance worse; and ii) the quasar luminosity function is more 
dominated by faint objects at low redshifts (see Fig. \ref{Fig:LumFun}), and 
since we have culled those objects with our absolute luminosity cut, $M_G<-22$, 
the volumetric density drops by a large factor, thus increasing shot noise.
}
\label{Fig:3zs_baos_errors}
\end{figure}

Figs. \ref{Fig:PkErrors}-\ref{Fig:3zs_baos_errors} demonstrate 
that quasars are not only viable tracers of large-scale structure,
but they can also detect the BAO features at high redshifts.
An interesting advantage of a high-redshift measurement of BAOs
is the milder influence of redshift distortions and nonlinear effects. 
In linear perturbation theory, the redshift-space and the real-space spectra 
are related by $P^{(s)}_q/P^{(r)}_q \simeq 1 + \frac23 \beta_q  + \frac15 \beta_q^2$
[\citet{kaiser_clustering_1987,hamilton_linear_1997}],
where $\beta_{q} \simeq \Omega_m^{0.55}/b_{q}$ in a flat
$\Lambda$CDM Universe.
Redshift distortions in the nonlinear regime are more difficult to take into account, but
they also scale roughly with $\beta_q$ -- see, e.g., 
\citet{jain_second-order_1994,meiksin_baryonic_1999,scoccimarro_power_1999,seo_high-precision_2009,smith_scale_2006,angulo_detectability_2007,seo_non-linear_2008}.
Since quasars become more highly biased at high redshifts, both linear and nonlinear
redshift-space distortions are suppressed relative to the local Universe.

The effect of random motions can be taken into accounted by 
multiplying the redshift-space spectrum factor of $1/[1+ k^2 \sigma_s(z)^2]$, 
where $\sigma_s(z)$ is a smoothing scale related to the one-dimensional 
pairwise velocity dispersion, and is usually calibrated by numerical simulations.
Nonlinear growth of structure and bulk flows (which tend to smear out the BAO signature)
also decrease at higher redshifts 
[\citet{smith_scale_2006,seo_non-linear_2008}].
\citet{angulo_detectability_2007}
found a useful parametrization of 
this effect in terms of a Fourier-space smoothing kernel 
$W(k,k_{nl})= \exp[-k^2/2k_{nl}^2]$, where $k_{nl}(z)$ is a non-linear
scale determined by numerical simulations.

\begin{figure}
\vspace{0.5cm}
\includegraphics[width=80mm]{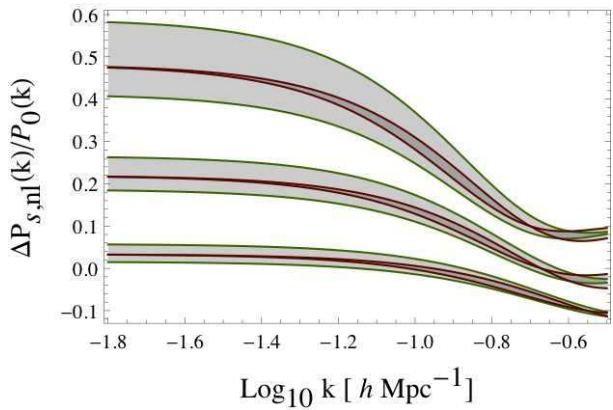}
\caption{
Scaling of the redshift distortions (outer, lighter contours and green
lines in color version) and of the effects of nonlinear structure
formation (inner, darker contours and red lines in color version),
for $z=1$, $z=2$ and $z=3$ from top to bottom, respectively. 
The uncertainties caused by redshift distortions and nonlinear effects, 
$\Delta_{s,nl}P/P_0$, are indicated by the hashed regions.
For visual clarity, we shifted the distortions at $z=1$ by $+0.1$, and 
the distortions at $z=3$ by $-0.1$.
We use the empirical calibration and errors of Angulo {\it et al.} 2008 for 
the redshift and nonlinear distortions. For the 
quasar bias and its uncertainties we employ the fit
of Ross {\it et al.} (2009).
}
\label{Fig:Nonlin}
\end{figure}

In Fig. \ref{Fig:Nonlin} we plot both the redshift distortions in linear
theory, and the nonlinear effects on the power spectrum.
For the redshift distortions we employ the quasar bias obtained
in \citet{ross_clustering_2009-1}:
$$
b_q = (0.53 \pm 0.19) + (0.289 \pm 0.035)(1+z)^2 \; ,
$$
which we assume holds up to $z=3$ (even though the uncertainties are
very large at such high redshifts.)
For the smoothing parameter we have extrapolated
the data from \citet{angulo_detectability_2007}, and found
$\sigma_s \simeq (4 - 0.96 z) h^{-1}$ Mpc (this approximation
is good up to $z\simeq3$.) 

Finally,
nonlinear structure formation effects are taken into account
by the nonlinear scale given in \citet{angulo_detectability_2007}
(which are appropriate for halos heavier than $M > 5 \times 10^{13} M_\odot$):
$$
k_{nl}(z) = (0.096 \pm 0.0074) + (0.036 \pm 0.0094) z \; ,
$$
in units of $h$ Mpc$^{-1}$.

With these assumptions, the ratio between the non-linear power spectrum in 
redshift space and the linear, position-space power spectrum is modeled by:
$$
\frac{P^{(s,nl)}_q (k,z)}{P^{(r,l)}_q(k,z)} = 1 + 
\left( \frac{1+\frac23 \beta + \frac15 \beta^2}{1+ k^2\sigma^2} -1 \right)
\, e^{ - k^2/2 k_{nl}^2} \; .
$$
Fig. \ref{Fig:Nonlin} illustrates that the distortions become 
smaller at higher redshifts, and that the 
uncertainties associated with them are also being suppressed.

In conclusion, we have seen that a large-area catalog of quasars, down to
depths of approximately $g<23$, can yield a precision measurement of the power
spectrum and of BAOs at moderate and high redshifts. 
The fact that quasars can measure large-scale structure
even better than LRGs around the peak of the power spectrum, despite
their much smaller volumetric density, can be understood as follows. First, the
volume spanned by quasars is larger, since they are much more luminous
and can be seen to higher redshifts than galaxies. This makes both 
sample variance {\it and} shot noise smaller by a factor of the square root of 
the volume, according to Eq. (\ref{Eq:StatErrPk}). Second, although the
number density of quasars is at least one order of magnitude smaller than
that of LRGs, the bias of quasars increases rapidly with redshift, and 
becomes higher than that of LRGs at $z\sim 1$. 
Since the volumetric factor which determines shot noise is the 
product of the number density and the square of the bias,
($n b^2$), a highly biased tracer such as quasars 
can afford to have a relatively small number density. At or near the peak of
the power spectrum, the accuracy of the power spectrum of quasars is 
almost limited by sample variance; slightly away from the peak, shot noise becomes
increasingly relevant, but the vast volume occupied by a catalog of quasars 
means that they are still superior compared to red galaxies. 
It is only on very small scales, where the amplitude of the power spectrum 
is very small, that galaxies become superior to quasars by virtue of their 
much higher number densities -- but then again, this only works at the relatively
low redshifts where galaxies can be efficiently observed.

How, then, could such a catalog of quasars be constructed? One possibility
is multi-object spectroscopy. 
While target selection of quasars from broad-band photometry can be
quite efficient in certain redshift ranges [such as z<2.2 for the SDSS
filter set [\citet{richards_photometric_2001}], there are ranges of redshifts where
the broad-band optical colors of quasars and the much more numerous
stars are indistinguishable, especially in the presence of photometric errors. 
There is the additional problem of contamination from galaxies,
but this should be a subdominant effect compared to stars (we leave this
issue for future work).
The comoving space density of quasars peaks between z=2.5 and
3, just the redshift at which the color locus of quasars crosses the
stellar locus [\citet{1999AJ....117.2528F}], and selecting quasars in this redshift range
tends to be quite inefficient and difficult [\citet{richards_efficient_2008,2011arXiv1105.0606R}].

A more concrete possibility is a narrow-band photometric survey, such as J-PAS, which will 
take low-resolution spectra of all objects (including quasars) in the surveyed 
area. However, there are two problems with this technique: first, unless 
the photometric redshifts of the quasars are very accurate, the relative errors in their radial
positions could be so large as to destroy their potential to map large-scale structure.
This is even more critical if we want to measure the signature of BAOs
in the angular and radial directions. In the previous Section we showed that it is
possible to obtain very small photo-z errors, so this should not be the main concern.
The main problem will be the classification of point sources as either quasars
or stars, especially for low-luminosity objects whose fluxes are noisy. Although
stars and quasars can be distinguished by criteria such as the $\chi^2$ of their
fits to templates, as well as the level of degeneracy of the PDF's of their photo-z's,
stars vastly outnumber quasars, and hence stringent criteria must be
used in order to preserve the purity of the quasar sample. This may compromise
the completeness (and therefore the final number density) of the quasar catalog, which
would then lead to large levels of shot noise. 
%This critical issue will be addressed in a forthcoming
%publication (Gonzalez-Serrano {\it et al.} 2012).

Hence, the key to realizing the potential of quasars to measure large-scale
structure in a narrow-band photometric survey hinges on whether or not we
can type a sufficiently high proportion of quasars, and obtain accurate 
photometric redshifts for the majority of objects in that catalog.
In the previous section we showed that this may be possible with 
an instrument such as J-PAS. However, our results can be easily generalized 
to other surveys such as Alhambra (which goes deeper than J-PAS, but has 
broader filters) and HETDEX (which subtends a smaller area and has a similar 
depth compared with J-PAS, but has much better spectral resolution).

% Even though shot noise is becoming increasingly important at
% higher redshifts, which can offset the suppression of
% redshift and nonlinearity distortions, the relative weights of the 
% systematic and statistical errors at high redshift will still be different 
% from those at low redshifts. The only way to forecast how these sources
% of uncertainty affect the estimation of the power spectrum, as
% well as the full power of the directional dependence, is through the Fisher
% matrix. We will leave a full Fisher matrix analysis for future work.

%\begin{figure}
%\vspace{0.5cm}
%\includegraphics[width=8.5cm]{Damp_factor_radial_BAOs_3ks}
%\caption{
%Damping of radial features on scales typical of the BAOs
%as a function of redshift, assuming that the all redshifts are
%used to measure the acoustic oscillations. This plot shows that
%it would be possible to make only a marginal measurement of
%radial BAOs with quasars.
%}
%\end{figure}

%%%%%%%%%%%%%%%%%%%%%%%%%%

\section{Conclusions}

We have argued that quasars are viable tracers of large-scale
structure in the Universe. A wide and deep survey of 
these objects will be a zero-cost consequence of several
ongoing or planned galaxy surveys that use either narrow-band filter systems
or integral field low-resolution spectroscopy.

Our estimates indicate that a dataset containing millions 
of objects will be a sub-product of these spectrophotometric 
surveys, and that they can lead not only to measurements
of the distribution of matter in the Universe up to very
high redshifts ($z \lesssim 4$), but also to an improved 
understanding of these objects, how they evolved,
what are their clustering properties and bias, as well as
their relationship and co-evolution with the host galaxies.
Such a large dataset, spread over such vast volumes, will
also allow a range of applications that break these objects 
into sub-groups (of absolute magnitude, types of host galaxies, etc.)

We have also shown that with a narrow-band set of filters
(of width $\sim$ 100 {\AA} in the optical) it is possible
to obtain near-spectroscopic photometric redshifts
for quasars: $\sigma_z \sim 0.001 (1+z)$ with the template
fitting method, and at least $\sigma_z \sim 0.002 (1+z)$ with the training set
method. This means an unprecedented resolution along the direction of 
the line-of-sight that extends up to vast distances, and is a further reason
for using quasars as tracers of large-scale structure.

\section*{Acknowledgements}
R. A. would like to thank N. Ben\'{\i}tez for stimulating discussions at
several stages of this work. Many thanks also to G. Bernstein, A. Bongiovanni,
I. Gonz\'alez-Serrano, B. Jain, A. Lidz, G. Richards, M. Sako, 
M. S\'anchez and R. Sheth for several enlightening discussions. 
R. A. would also like to thank the Department of
Astrophysical Sciences at Princeton University, as well as
the Department of Physics and Astronomy at the University of
Pennsylvannia, for their hospitality during the period when
this work was done.
MAS acknowledges the support of NSF grant AST-0707266.
This work was also partially supported by FAPESP and
CNPq of Brazil.

\bibliographystyle{mn2e.bst}
%\bibliography{/Users/lrwa/trabalho/abramo}

\end{document}